\documentclass[a4paper,11pt]{article}
\pdfoutput=1 

\usepackage{jcappub} 
\usepackage{amsmath}
\usepackage{booktabs}
\usepackage{threeparttable}
\usepackage{tabularx}

\usepackage{listings}

\usepackage[T1]{fontenc} 
\newcommand{\ion}[2]{\mathrm{#1\,\textsc{\lowercase{#2}}}}

\title{LIMFAST. IV. Learning High-Redshift Galaxy Formation from Multiline Intensity Mapping with Implicit Likelihood Inference}

\author[a,b,*]{Guochao Sun,\note[*]{Corresponding author.}}
\author[a,b]{Tri Nguyen,}
\author[a,b]{Claude-Andr\'{e} Faucher-Gigu\`{e}re,}
\author[c]{Adam Lidz,}
\author[a,b]{Tjitske Starkenburg,}
\author[a,b]{Bryan R. Scott,}
\author[d,e]{Tzu-Ching Chang,}
\author[f]{and Steven R. Furlanetto \vspace{0.15cm}}

\affiliation[a]{\textit{CIERA and Department of Physics and Astronomy, Northwestern University, 1800 Sherman Ave, Evanston, IL 60201, USA}}
\affiliation[b]{\textit{NSF--Simons AI Institute for the Sky (SkAI), 172 E. Chestnut St., Chicago, IL 60611, USA}}
\affiliation[c]{\textit{University of Pennsylvania, Department of Physics and Astronomy, 209 S. 33rd Street, Philadelphia, PA 19104, USA}}
\affiliation[d]{\textit{Jet Propulsion Laboratory, California Institute of Technology, 4800 Oak Grove Drive, Pasadena, CA 91109, USA}}
\affiliation[e]{\textit{California Institute of Technology, 1200 E. California Blvd., Pasadena, CA 91125, USA}}
\affiliation[f]{\textit{Department of Physics and Astronomy, University of California, Los Angeles, CA 90095, USA \vspace{0.15cm}}}

\emailAdd{guochao.sun@northwestern.edu}
\emailAdd{trivtnguyen@northwestern.edu}
\emailAdd{cgiguere@northwestern.edu}
\emailAdd{alidz@sas.upenn.edu}
\emailAdd{tjitske.starkenburg@northwestern.edu}
\emailAdd{bryan.scott@northwestern.edu}
\emailAdd{tzu-ching.chang@jpl.nasa.gov}
\emailAdd{sfurlane@astro.ucla.edu}

\abstract{
By opening up new avenues to statistically constrain astrophysics and cosmology with large-scale structure observations, the line intensity mapping (LIM) technique calls for novel tools for efficient forward modeling and inference. Implicit likelihood inference (ILI) from semi-numerical simulations provides a powerful setup for investigating a large model parameter space in a data-driven manner, therefore gaining significant recent attention. Using simulations of high-redshift 158\,$\mu$m [$\ion{C}{II}$] and 88\,$\mu$m [$\ion{O}{III}$] LIM signals created by the \texttt{LIMFAST} code, we develop an ILI framework in a case study of learning the physics of early galaxy formation from the auto-power spectra of these lines or their cross-correlation with galaxy surveys. We leverage neural density estimation with normalizing flows to learn the mapping between the simulated power spectra and parameters that characterize the physics governing the star formation efficiency and the $\dot{\Sigma}_{\star}$--$\Sigma_\mathrm{g}$ relation of high-redshift galaxies. Our results show that their partially degenerate effects can be unambiguously constrained when combining [$\ion{C}{II}$] with [$\ion{O}{III}$] measurements to be made by new-generation mm/sub-mm LIM experiments. 
}


\begin{document}

\maketitle
\flushbottom


\section{Introduction}
\label{sec:intro}

Deep, detailed observations of an increasingly large sample of high-redshift ($z \gtrsim 6$) galaxies made by new-generation telescopes like the James Webb Space Telescope (JWST) have transformed the understanding of physical processes that govern galaxy formation in the early universe \citep{Robertson2022,Adamo2024arXiv}. A major constraint for such flux-limited observations, however, is that only galaxies brighter than a certain survey depth can be detected within a given area. This observational trade-off between survey depth and area restricts studies of individual galaxies to a limited sample of relatively bright sources, potentially leaving a significant population of fainter sources undetected. While their exact contribution to cosmic reionization remains debated \citep{KuhlenFaucherGiguere2012,Robertson2015,Naidu2020,Munoz2024}, the population of faint galaxies below JWST's typical detection threshold is predicted to be responsible for more than half of the total cosmic SFRD at $z>10$ \citep{Furlanetto2017}. Meanwhile, sensitive to the fundamentals of galaxy formation and evolution, especially the roles of stellar feedback and dark matter \citep{FerraraTolstoy2000,Mashchenko2008,Wetzel2016}, faint/low-mass galaxies provide crucial insights into early structure formation and are therefore of special importance observationally.

Line intensity mapping (LIM) promises to complement galaxy surveys by measuring the spatial fluctuations in the aggregate emission of certain lines (e.g., 21\,cm, Ly$\alpha$, and [$\ion{C}{II}$]) sourced by the entire galaxy population, including the very faint ones for which individual detections are unlikely \citep{Visbal2010,BernalKovetz2022}. Numerous studies have employed empirical/semi-analytic models or simulations to demonstrate that rich physical information about high-$z$ galaxy formation may be revealed by LIM surveys of multiple tracers \citep[see e.g.,][]{Sun2019,SchaanWhite2021a,SchaanWhite2021b,Yang2021,Kannan2022,Parsons2022,Sun2022,LIMFAST1,LIMFAST2,Roy2023,Sato-Polito2023,Libanore2025}. Multi-tracer analysis based on cross-correlations also benefits from its lower susceptibility to observational biases due to foreground contamination \citep{Lidz2011,Beane2019,RoyBattaglia2024}. In the context of probing the galaxy population responsible for cosmic reionization, LIM observations of the 158\,$\mu$m [$\ion{C}{II}$] and 88\,$\mu$m [$\ion{O}{III}$] lines (redshifted into mm/sub-mm wavelengths) are promising. Not only is this because they are the two brightest far-infrared emission lines from high-$z$ galaxies that have been routinely detected at increasing redshift to provide precise spectroscopic redshifts \citep{Harikane2020,Bakx2023,Heintz2023}, but they also encode rich physical information about the stellar population and interstellar medium (ISM) in the infant universe \citep{Algera2024,Schimek2024}. Padmanabhan et al. (2022) \citep{Padmanabhan2022}, in particular, investigate the potential for synergizing [$\ion{C}{II}$] and [$\ion{O}{III}$] LIM at the reionization era across a variety of experimental designs and find that improvements to current-generation mm/sub-mm LIM experiments could enable such synergy in the foreseeable future. 

With the advent of new LIM experiments that will enable joint analysis of multiple lines, an urgent need is created for statistical inference tools to constrain specific physical processes governing galaxy formation with LIM data. Bayesian sampling of the multidimensional posterior distribution---assuming some explicit likelihood constructed from analytically modeled summary statistics---is the standard approach in traditional cosmological inference. It has been extensively applied to parameter inference from $\ion{H}{I}$ 21\,cm LIM simulations \citep{GreigMesinger2015,Park2019}, though the high dimensional and weakly constrained parameter space often makes on-the-fly sampling with methods such as Markov Chain Monte Carlo (MCMC) computationally expensive \citep{Kern2017,Maity2023,Mason2023}. The explicit likelihood can not only lead to high computational cost as individual evaluations add up, but also introduce unwanted systematic bias in the inferred posterior due to strong assumptions made about e.g., the noise distribution. 

Implicit likelihood inference (ILI), also known as simulation-based inference, provides a powerful and practical solution to these limitations through approaches such as Neural Posterior Estimation (NPE) and has therefore gained increasing popularity \citep{Cranmer2020PNAS}, with applications in astrophysics and cosmology expanding significantly in recent years \citep[e.g.,][]{Alsing2019,Dax2021,Wang2023,Nguyen2023,Ho2024,Modi2025,Nguyen2025}. With techniques such as normalizing flows trained on simulated data, ILI directly constructs the posterior distribution using neural networks without assuming an explicit likelihood function. This enables efficient analysis of complex, high-dimensional parameter spaces, with scalability and flexibility stemming from the expressive power and computational efficiency of neural network-based architectures. Several recent studies have demonstrated the potential of ILI as an efficient and reliable method for inferring reionization parameters from the 21\,cm LIM signal \citep[e.g.,][]{Zhao2022LC, Zhao2022PS, PrelogovicMesinger2023,Saxena2023}. As LIM of other emission lines beyond $\ion{H}{I}$ 21\,cm, especially those correlated with the galaxy star formation rate (SFR) such as [$\ion{C}{II}$] and [$\ion{O}{III}$], also shows significant promise for probing high-$z$ galaxy formation, it is likewise interesting to consider ILI for efficient Bayesian parameter inference from those lines, including their synergies with other large-scale structure tracers like the distribution of galaxies. 

In this paper, we study the constraints on high-$z$ galaxy formation physics from [$\ion{C}{II}$] and [$\ion{O}{III}$] LIM measurements using an ILI framework. As a proof-of-concept case study, we focus on both their respective auto-correlations and their cross-correlation with galaxy redshift surveys by the Roman Space Telescope, which can place stringent constraints on physical processes that determine the star formation efficiency (SFE) and gas content of high-$z$ galaxies. For the first time to our best knowledge, we establish an analysis framework to explicitly demonstrate the power of multi-tracer LIM for constraining a broad class of simple physical models describing feedback-regulated star formation in high-$z$ galaxies. We base our analysis on the training and testing data created by the semi-numerical \texttt{LIMFAST} simulations \citep{LIMFAST1,LIMFAST2,LIMFAST3}, covering a reasonably diverse range of model variations. Taking specifications of new-generation mm/sub-mm LIM experiments, such as FYST/CCAT-prime \citep{CCAT-Prime2023} and TIFUUN \citep{Kohno2024SPIE}, along with their hypothesized successors, that target these lines emitted by high-$z$ galaxies, we further predict the constraining power these experiments may provide on the physical parameters of interest. 

We organize the remainder of this paper as follows. In section~\ref{sec:models}, we summarize the key steps to simulate the [$\ion{C}{II}$] and [$\ion{O}{III}$] LIM signals of interest in \texttt{LIMFAST} based on an analytic galaxy formation model. We describe how we develop the ILI framework in section~\ref{sec:ili} and present the main results on the predicted parameter constraints in section~\ref{sec:results}. We discuss some noteworthy implications and possible extensions of the current framework, before concluding in section~\ref{sec:discussion}. A flat, $\Lambda$CDM cosmology consistent with measurements by Planck Collaboration et al. (2016) \citep{Planck2016} is adopted.  


\section{Simulating $[\ion{C}{II}]$ and $[\ion{O}{III}]$ LIM Signals} \label{sec:models}

Semi-numerical simulations offer a fast and flexible way to model large-scale structure formation and the resulting radiation fields. As will be detailed in this section, in order to create the LIM data required to build our ILI framework, we adopt the models and features previously introduced into \texttt{LIMFAST}, which itself is an extension of the \texttt{21cmFAST} code \citep{Mesinger2011,Park2019} widely used for 21\,cm cosmology in the eras of cosmic dawn and reionization. We simulate at discrete redshifts two important summary statistics of [$\ion{C}{II}$] and [$\ion{O}{III}$] lines, namely the auto-correlation power spectrum and the cross-correlation power spectrum with discrete high-$z$ galaxies. Observational uncertainties are modeled using realistic specifications of the instrument and survey design.  

\subsection{Galaxy formation and line emission models}

\begin{figure*}
 \centering
 \includegraphics[width=\textwidth]{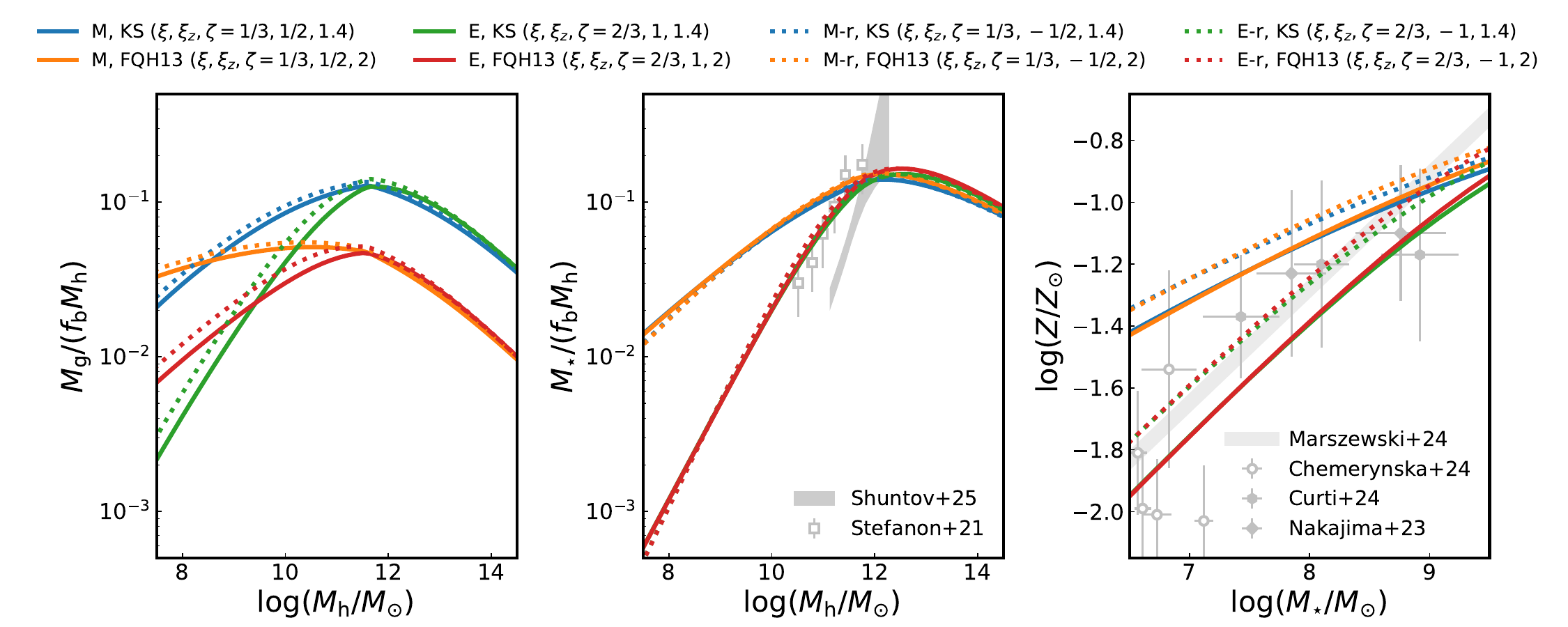}
 \caption{The predicted gas mass--halo mass relation (left), stellar mass--halo mass relation (middle), and mass-metallicity relation (MZR, right) at $z=6$ under varying assumptions of the mass loading factor ($\xi$ and $\xi_z$) and $\dot{\Sigma}_{\star}$--$\Sigma_\mathrm{g}$ relation ($\zeta$) parameters. Note that the stellar mass content is set by the feedback strength, e.g., momentum-driven (``M'') or energy-driven (``E''), and largely independent of the $\dot{\Sigma}_{\star}$--$\Sigma_\mathrm{g}$ relation (``KS'' or ``FQH13''), which primarily affects the gas mass content. The insensitivity of the MZR to the $\dot{\Sigma}_{\star}$--$\Sigma_\mathrm{g}$ relation is then a necessary consequence of the metal mass production at equilibrium. As sanity checks, we plot observational constraints on the star formation efficiency at $z\sim6$ from HST \citep{Stefanon2021} and JWST \citep{Shuntov2025}, along with the high-$z$ MZR predicted by the FIRE simulations \citep{Marszewski2024}, which is in close agreement with the latest JWST observations \citep[e.g.,][]{Nakajima2023,Chemerynska2024,Curti2024}.}
 \label{fig:halo_properties}
\end{figure*}

In this work, we take the same physical prescriptions of galaxy formation and line emission as described in Sun et al. (2023) \citep{LIMFAST2}. Simple, physically motivated models are often used to interpret individually detected high-$z$ galaxy samples \citep[e.g.,][]{SF2016,Tacchella2018,SippleLidz2024,Donnan2025}, and adopting a model of similar kind here helps illustrate the complementary constraints provided by LIM measurements. A notable modification, however, is that we introduce more explicit power-law parameterizations of the mass loading factor regulating the SFE\footnote{In this work, we define the SFE to be the ratio of the SFR to the baryonic mass accretion rate of a halo, as prescribed by the solution to the ODEs.} and the $\dot{\Sigma}_{\star}$--$\Sigma_\mathrm{g}$ relation (sometimes referred to as the ``star formation law'') for the convenience of model inference. Specifically, following Furlanetto (2021) \citep{Furlanetto2021}, we define the mass loading factor $\eta$, which specifies the strength of stellar feedback that regulates the SFE, and the $\dot{\Sigma}_{\star}$--$\Sigma_\mathrm{g}$ relation, namely the relation between the SFR and gas surface densities that characterizes the gas depletion time at different densities, as, 
\begin{equation}
\eta(M_\mathrm{h}, z) = \eta_0 \left( \frac{M_\mathrm{h}}{10^{11.5}\,M_{\odot}} \right)^{-\xi} \left( \frac{1+z}{9} \right)^{-\xi_z},
\end{equation}
and
\begin{equation}
\dot{\Sigma}_{\star}(\Sigma_\mathrm{g}, z) = \epsilon_{\star,0} \frac{\Sigma_\mathrm{g,0}}{t^\mathrm{disc}_\mathrm{ff,0}} \left( \frac{\Sigma_\mathrm{g}}{\Sigma_\mathrm{g,0}} \right)^{\zeta} \left( \frac{1+z}{9} \right)^{\zeta_z},
\end{equation}
where the normalization factors ($\eta_0 = 2.5$, $\epsilon_{\star,0}=0.015$, $\Sigma_\mathrm{g,0} = 100\,M_{\odot}\,\mathrm{pc}^{-2}$, and $t^\mathrm{disc}_\mathrm{ff,0}=2\,$Myr) are set such that the predicted galaxy number statistics and scaling relations are broadly consistent with observations (see figure~\ref{fig:halo_properties}). The power-law indices are left as free parameters, which, along with other physical quantities such as the halo mass threshold, $M_\mathrm{min}$, below which the contribution to the total cosmic star formation is negligible\footnote{This mass scale, sometimes referred to as the minimum halo mass for star formation, depends on the ability of accreted gas to cool, the impact of external feedback (e.g., photoionization and photoheating), and the stellar population concerned (e.g., Population~II vs. Population~III stars). While not treated as a free parameter, it can be degenerate with parameters like $\xi$ and should be considered with caution in more complete analyses.}, affect the shape and amplitude of LIM summary statistics and thus can be constrained through Bayesian inference. To focus our model inference analysis on the SFE and the $\dot{\Sigma}_{\star}$--$\Sigma_\mathrm{g}$ relation, we assume a redshift-specific value for $M_\mathrm{min}$ that corresponds to a constant virial temperature of $T_\mathrm{vir} = 10^4\,$K (temperature threshold for efficient atomic cooling via collisionally excited H and He lines). 

Assuming that the stellar and gas mass contents of galaxy host halos are in a quasi-equilibrium state maintained by star formation and feedback, a system of ordinary differential equations (ODEs) defines a simple, physical framework that specifies key galaxy properties from which halo emissivities of different lines can be calculated. Defining $M^{\prime} \equiv d M / d z$ and $\tilde{M} \equiv M / M_\mathrm{h,0}$ and writing the evolution of halo properties in terms of surface densities $\Sigma \equiv M / (2\pi R^2_{1/2})$ with the half-mass disc radius $R_{1/2} \approx 0.02R_\mathrm{vir}$ (assumed to be identical for gas and stars), we can compute the redshift evolution of non-dimensionalized halo properties, including the halo mass ($M_\mathrm{h}$), gas mass ($M_\mathrm{g}$), stellar mass ($M_\mathrm{\star}$), and metal mass ($M_{Z}$), using the following system of ODEs \citep{Furlanetto2021},
\begin{equation}
\frac{\tilde{M}^{\prime}_\mathrm{h}}{\tilde{M}_\mathrm{h}} = -\mathcal{M}_0, 
\end{equation}
\begin{equation}
\frac{\tilde{M}^{\prime}_\mathrm{g}}{\tilde{M}_\mathrm{g}} = \mathcal{M}_0 \left[ -\frac{1}{X_\mathrm{g}} + \eta_0 \dot{X}_{\star,0} \left( \frac{X_\mathrm{g}}{X_\mathrm{g,0}} \right)^{\alpha_X}  \tilde{M}^{\alpha_m}_\mathrm{h} \left( \frac{1+z}{1+z_0} \right)^{\alpha_z} \right], 
\end{equation}
\begin{equation}
\tilde{M}^{\prime}_{\star} = -\mathcal{M}_0 \dot{X}_{\star,0} \left( \frac{X_\mathrm{g}}{X_\mathrm{g,0}} \right)^{\beta_X} \tilde{M}^{\beta_m}_\mathrm{h} \left( \frac{1+z}{1+z_0} \right)^{\beta_z}, 
\end{equation}
\begin{equation}
\frac{X^{\prime}_\mathrm{g}}{X_\mathrm{g}} =  \frac{\tilde{M}^{\prime}_\mathrm{g}}{\tilde{M}_\mathrm{g}} - \frac{\tilde{M}^{\prime}_\mathrm{h}}{\tilde{M}_\mathrm{h}},
\end{equation}
\begin{equation}
\tilde{M}'_{Z} = \left[ y_Z - \eta \left( \tilde{M}_{Z} / \tilde{M}_\mathrm{g} \right) \right] \tilde{M}^{\prime}_{\star},
\label{eq:metal}
\end{equation}
where $\mathcal{M}_0 = \dot{M}_\mathrm{h,0}/M_\mathrm{h,0}/H(z_0)/(1+z_0)$ and the gas retention factor $X_\mathrm{g} = M_\mathrm{g}/(f_\mathrm{b} M_\mathrm{h})$ with $f_\mathrm{b} = 0.16$. In the limit $\eta \gg 1$, power-law indices for the gas mass and stellar mass evolution can be written as $\alpha_{X} = \zeta - 1, \alpha_{m} = (\zeta - 3\xi - 1)/3, \alpha_{z} = 2\zeta - 4.5 + \zeta_z, \beta = \zeta, \beta_m = (2+\zeta)/3$, and $\beta_z = 2\zeta - 4.5 + \zeta_z$. At high halo masses where $\eta \gg 1$ no longer holds, stellar feedback can become less dominant compared to other quenching mechanisms such as AGN feedback. Here, as in Furlanetto et al. (2017) \citep{Furlanetto2017}, we adopt a simple treatment that effectively suppresses star formation in massive halos by a factor \citep[][]{CAFG2011}\footnote{It is worth noting that although Equation~(\ref{eq:fs}) was originally introduced to describe how heating by virial shocks affects the fraction of accreted gas that can cool onto the galaxy, shock heating alone is insufficient to quench star formation in massive halos. Rather, the presence of hot, virialized circumgalactic gas establishes the conditions necessary for the efficient launch and coupling of AGN feedback that suppresses star formation \citep[][]{DekelBirnboim2006,Keres2009,Stern2020,Byrne2024}. Nevertheless, for the purpose of this work, Equation~(\ref{eq:fs}) serves as a reasonable approximation for star formation quenching in massive halos.}
\begin{equation}
f_\mathrm{s} = \mathrm{min}\left[ 0.47 \left( \frac{M_\mathrm{h}}{10^{12}\,M_{\odot}} \right)^{-1/4} \left( \frac{1+z}{4} \right)^{0.38}, 1 \right],
\label{eq:fs}
\end{equation}
such that the effective mass loading factor is $\eta_\mathrm{eff} \simeq \eta / f_\mathrm{s}$. 

We note that Equation~(\ref{eq:metal}) for the metal mass evolution is highly simplified, involving only one source term for metal enrichment by star formation and one sink term for metal ejection due to feedback-driven outflows. More generally, the evolution of high-$z$ galaxies is in reality governed by far more complex physics than what is naively described by the system of ODEs considered here. Nevertheless, this framework provides physically grounded approximations to the observed scaling relations of halo properties despite its simplicity. Our galaxy formation model should therefore be regarded as a minimal representation of the relevant physics, which is sufficient for proof-of-concept studies of LIM observables and serves as a foundation for future extensions with more accurate physical models beyond the simple ODEs. For example, analytic descriptions of the bursty and turbulence-dominated nature of galaxies in their early stage of formation \citep{FurlanettoMirocha2022,HegdeFurlanetto2025,Pallottini2025,Sun2025}, motivated by both observations and numerical simulations, could be incorporated to more realistically model the high-$z$ galaxy formation history compared with the equilibrium solutions considered here. 

In figure~\ref{fig:halo_properties}, we show the predicted scaling relations of the gas mass, stellar mass, and metallicity of halos at $z=6$ and how they depend on the detailed physical prescriptions in our galaxy formation model. Eight model variations characterized by different combinations of power-law indices $\xi$, $\xi_z$, and $\zeta$ are shown as examples, which correspond to momentum-driven (``M'') vs. energy-driven (``E'') stellar feedback (with/without the redshift dependence reversed) and the $\dot{\Sigma}_{\star}$--$\Sigma_\mathrm{g}$ relation from \citep[][]{Kennicutt1998} (``KS'') and \citep[][]{CAFG2013} (``FQH13''). The coupling mechanism of stellar feedback determines how the mass loading factor depends on $M_\mathrm{h}$ and $z$ (the latter results from the definition of $M_\mathrm{h}$ relative to the mean/critical density of the Universe), with $\xi = 1/3$ (2/3) and $\xi_z = 1/2$ (1) corresponding to the case where the feedback momentum (energy) is conserved. The KS $\dot{\Sigma}_{\star}$--$\Sigma_\mathrm{g}$ relation has a slope of $\zeta = 1.4$ consistent with disk-averaged measurements of local star-forming galaxies, whereas the FQH13 $\dot{\Sigma}_{\star}$--$\Sigma_\mathrm{g}$ relation with a steeper slope of $\zeta = 2$ is derived for a turbulent gas disk whose vertical support is provided by stellar feedback. 

For context, we include constraints on the stellar mass and metal contents of halos obtained by HST and JWST observations. The broad agreement between our model predictions and the observed constraints suggests that, despite its simplicity, the galaxy formation model considered here successfully captures some of the key physics that can be inferred from observations. From the comparison of model variations, the halo gas mass content is significantly affected by both stellar feedback (through $\xi$ and $\xi_z$) and the $\dot{\Sigma}_{\star}$--$\Sigma_\mathrm{g}$ relation (through $\zeta$), whereas the stellar mass content or SFE is barely affected by the $\dot{\Sigma}_{\star}$--$\Sigma_\mathrm{g}$ relation. These are well-known consequences of self-regulated star formation: the gas content is able to adjust itself so that ``balanced'' amounts of star formation and stellar feedback are produced. The strongly diverging behaviors at low masses imply that LIM measurements as integral constraints particularly sensitive to low-mass galaxy populations can have strong constraining power on these model variations. 

Joint [$\ion{C}{II}$] and [$\ion{O}{III}$] LIM observations have been proposed as a promising way to trace high-$z$ galaxies during cosmic reionization by \cite{Padmanabhan2022}, although the nature of their complementarity remains unexplored. With the derived halo properties in hand, we employ the same \texttt{Cloudy} \cite{Ferland2017}-based gas nebula model introduced in \cite{LIMFAST1} and \cite{LIMFAST2} to compute the halo emissivities of [$\ion{C}{II}$] and [$\ion{O}{III}$] lines, which then allow \texttt{LIMFAST} to generate the LIM signals using the evolved density field. Briefly, the gas nebula model relies on pre-calculated lookup tables evaluated for \texttt{Cloudy} (v17.03) model grids spanning a range of ISM conditions (e.g., density, metallicity, ionization parameter), along with a subgrid prescription for the gas density distribution for turbulent molecular clouds in the ISM following \citep{Vallini2018}. Interested readers are referred to \cite{LIMFAST2} for further details of the line emission modeling in \texttt{LIMFAST}. 

Originating mainly from the cooling process balancing photoelectric heating in photo-dissociation regions (PDRs) of the ISM, the 158\,$\mu$m [$\ion{C}{II}$] line depends on galaxy properties such as the gas mass and metallicity, as well as other ISM conditions like the interstellar radiation field \citep{Lagache2018,Ferrara2019}. Although frequently considered as an SFR tracer given the tight, nearly redshift-independent $L_\mathrm{[CII]}$--SFR correlation \citep{Schaerer2020,Sun2021}, $L_\mathrm{[CII]}$ is not simply governed by the SFR and is physically more directly tied to the halo gas content. On the other hand, thanks to the ionization potential of $\ion{O}{II}$ (35.1\,eV), the 88\,$\mu$m [$\ion{O}{III}$] line originates solely from the HII regions and strongly depends on the photoionization rate, which in turn scales with the formation rate of short-lived O \& B stars \citep{YangLidz2020}. Both lines are also sensitive to other ISM conditions, such as metallicity and the ionization parameter, whose effects are captured in \texttt{Cloudy} models. 

\begin{figure}
 \centering
 \includegraphics[width=\columnwidth]{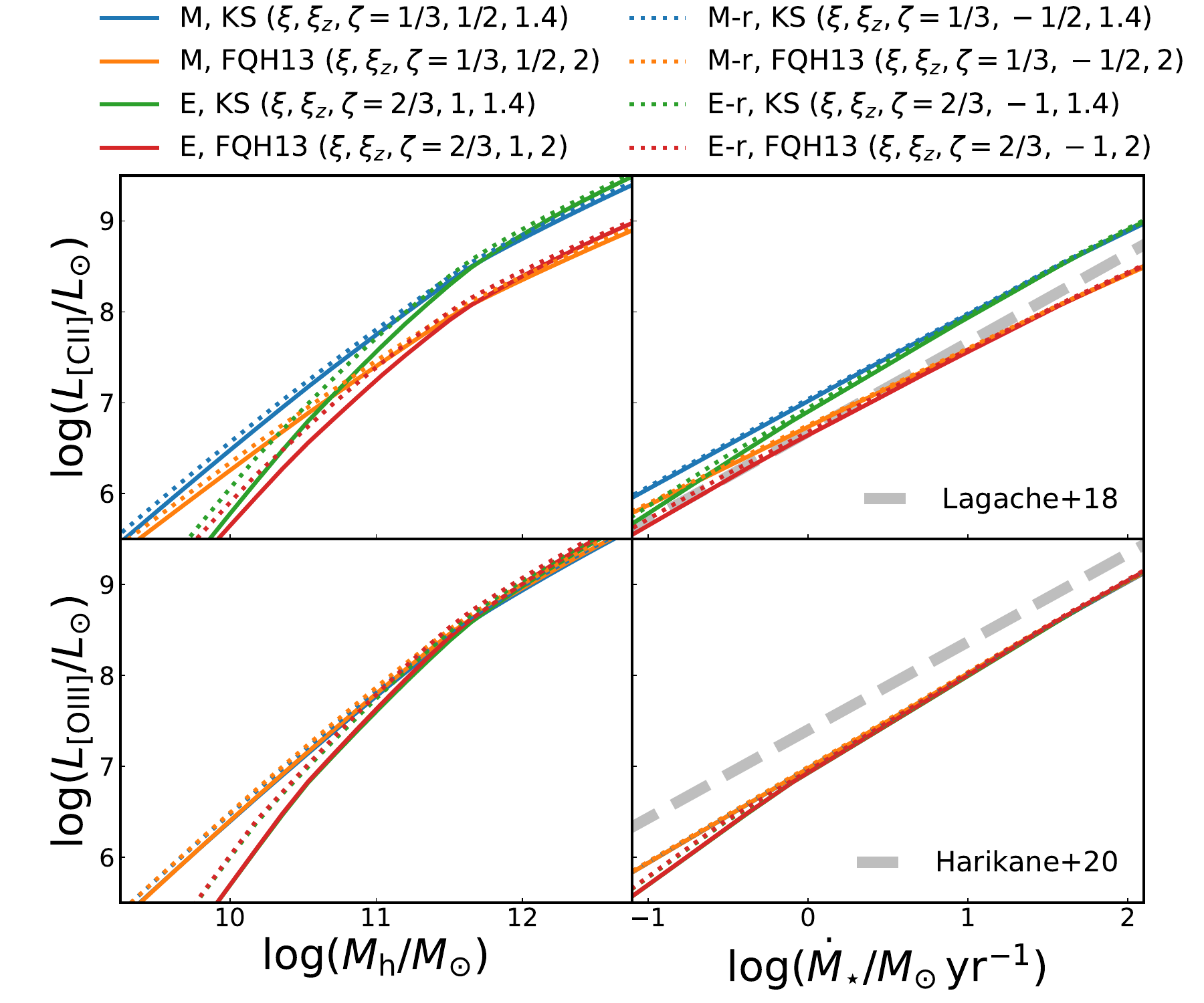}
 \caption{Scaling relations of [$\ion{C}{II}$] and [$\ion{O}{III}$] luminosities as a function of the halo mass (left) or the SFR (right) at $z = 6$ for the same model variations as in figure~\ref{fig:halo_properties}. Note how the [$\ion{C}{II}$] luminosity depends on both the feedback mode and the $\dot{\Sigma}_{\star}$--$\Sigma_\mathrm{g}$ relation as a result of their effects on the gas mass, whereas the [$\ion{O}{III}$] luminosity is almost independent of the latter due to the lack of sensitivity of the SFR to $\zeta$. For comparison, the predictions of some other models in the literature are shown by the gray dashed curves, which provide good fits to the latest ALMA observations \citep{Lagache2018,Harikane2020}.}
 \label{fig:halo_luminosities}
\end{figure}

In figure~\ref{fig:halo_luminosities}, we show the predicted luminosity--halo mass and luminosity--SFR relations at $z=6$ for the same model variations as considered in figure~\ref{fig:halo_properties}, in comparison with the scaling relations in the literature that agree with the latest ALMA observations \cite{Lagache2018,Harikane2020}. As expected, because $L_\mathrm{[CII]}$ closely traces the gas content whereas $L_\mathrm{[OIII]}$ roughly scales with the amount of star formation, the mass and redshift dependence of line luminosities generally reflects the underlying mass and redshift dependence of the corresponding halo properties illustrated in figure~\ref{fig:halo_properties}. Since the luminosity--halo mass relation can be sensitively probed by LIM summary statistics like the power spectrum \cite{Visbal2010,SchaanWhite2021a}, it is reasonable to expect that the degenerate effects of stellar feedback and the $\dot{\Sigma}_{\star}$--$\Sigma_\mathrm{g}$ relation on [$\ion{C}{II}$] LIM observables may be better distinguished when combined with [$\ion{O}{III}$] constraints. 

\begin{figure*}
 \centering
 \includegraphics[width=\textwidth]{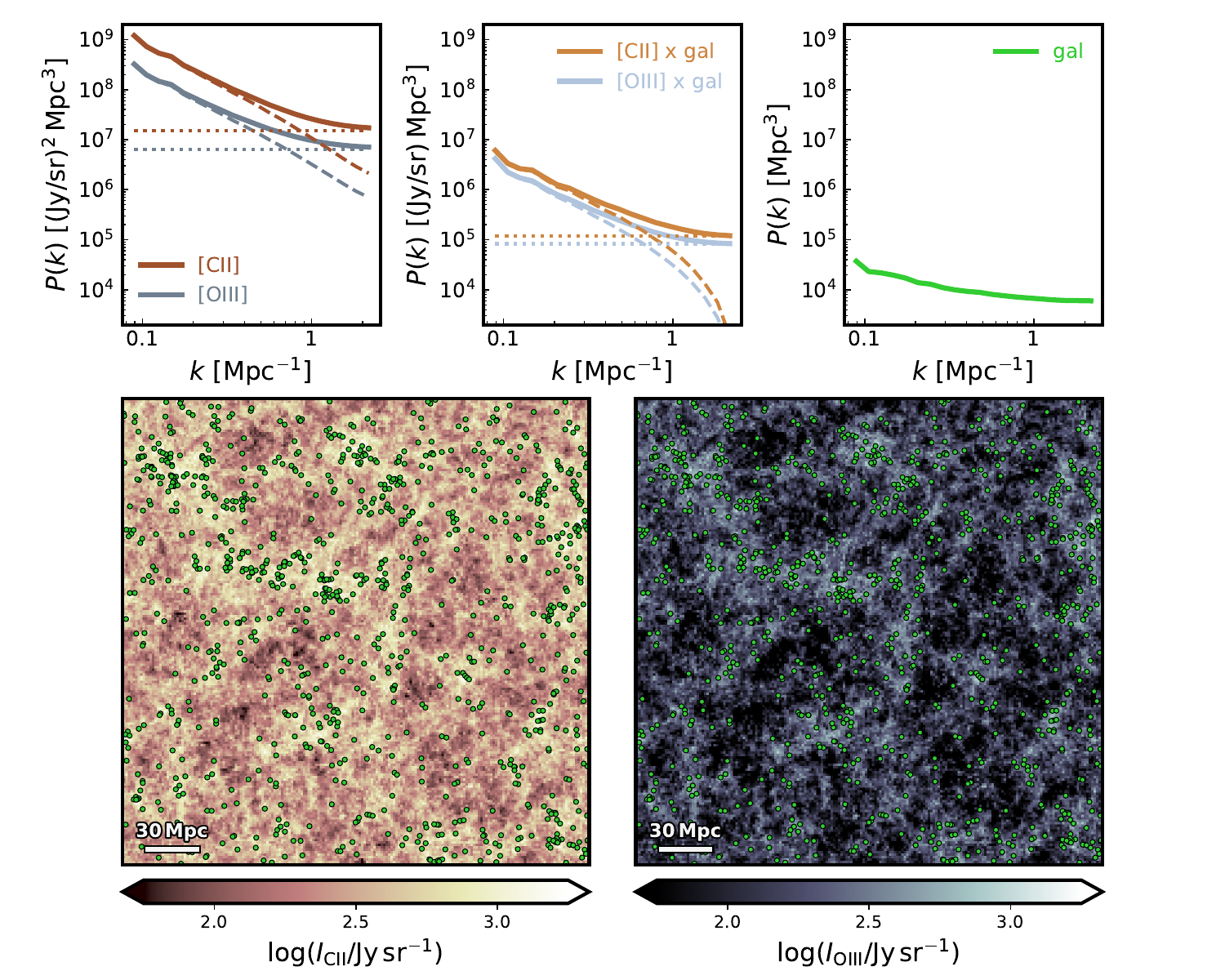}
 \caption{Top: shot-noise-included power spectra of [$\ion{C}{II}$] and [$\ion{O}{III}$] (left), their cross-correlations with the Roman LBGs (middle), and the LBGs themselves (right) at $z = 7$ for an example model with $\xi \approx 1/3$, $\xi_z \approx 0$, and $\zeta \approx 1.4$. The clustering and shot-noise components of [$\ion{C}{II}$] and [$\ion{O}{III}$] power spectra are indicated by the dashed and dotted curves, respectively. Bottom: mock [$\ion{C}{II}$] and [$\ion{O}{III}$] intensity maps at $z = 7$. Locations of LBGs detectable for a moderate depth survey by the Roman ($m_\mathrm{AB, lim} < 28.2$) are marked by the green dots, which correlate with the overdensities traced by the line intensity maps. }
 \label{fig:rendering}
\end{figure*}

\subsection{LIM observables}

For the purpose of this paper, we use \texttt{LIMFAST} to simulate co-eval boxes of [$\ion{C}{II}$] and [$\ion{O}{III}$] intensities with dimension 256$^3$\,cMpc$^3$ and resolution 1.5$^3$\,cMpc$^3$, which are chosen to roughly match the survey and instrument specifications considered in table~\ref{tb:sens}. We then consider LIM observables in two cases. In the first case, we focus on redshifts $z = 6$--9 for [$\ion{C}{II}$] and on $z = 7.5$ and 8.5 for [$\ion{O}{III}$], motivated by the mm/sub-mm atmospheric transmission windows in which these two lines can be accessed by ground-based experiments (e.g., FYST/CCAT-prime and TIFUUN). Other specifications, including the level of noise power $P_\mathrm{N}$, are also chosen to be comparable to forthcoming surveys by FYST/CCAT-prime and TIFUUN. In the second case, we consider a hypothetical space-based experiment targeting both [$\ion{C}{II}$] and [$\ion{O}{III}$] over $z=6$--9 with lower $\sigma_\mathrm{N}$ values from space, similar to the design proposed in table~2 of \cite{Padmanabhan2022}. While in this work we do not attempt to make detailed forecasts for specific experiments, these two cases detailed in table~\ref{tb:sens} are realistic representations of the capabilities of current-generation, ground-based facilities and terahertz space missions for joint [$\ion{C}{II}$] and [$\ion{O}{III}$] LIM in the foreseeable future. 

\subsubsection{$[\ion{C}{II}]$ and $[\ion{O}{III}]$ auto-power spectra}

The auto-correlation power spectrum is the most commonly used summary statistic of LIM data that informs about the source population through spatial fluctuations of the line intensity. For the purpose of our analysis in this work, we compute auto-power spectra, $P_{I}$, directly from the coeval, signal-only line intensity boxes generated by our simulations, 
\begin{equation}
\langle \tilde{I}(\boldsymbol{k}, z) \tilde{I}^{*}(\boldsymbol{k}', z) \rangle = (2\pi)^3 P_I(\boldsymbol{k}, z) \delta^{3}_{D}(\boldsymbol{k}-\boldsymbol{k}'),
\end{equation}
where $\tilde{I}(\boldsymbol{k}, z)$ is the Fourier transform of the line intensity field, $I(\boldsymbol{x}, z)$, and $\delta^{3}_{D}$ is the 3D Dirac delta function. In addition to the contribution from two-point clustering, since individual, bright [$\ion{C}{II}$] and [$\ion{O}{III}$] emitters are not resolved in our semi-numerical simulations, we supplement the directly computed auto-power spectra with a scale-independent shot-noise component\footnote{We note that the simplified treatment of shot noise adopted here comes with a few caveats. First, deviations from the simple Poissonian approximation are expected on small scales due to halo and galaxy stochasticity (see e.g., \citep{Baldauf2013,Jun2025}). In addition, even with matched cosmological parameters, halo mass function, etc., small inconsistencies may still exist between the analytically modeled shot noise and the simulated spatial fluctuations on larger scales, which can bias the inference. Future simulations that resolve the entire source population will enable more accurate modeling across relevant scales and help improve the robustness of ILI analysis.} evaluated \textit{analytically} by the integral
\begin{equation}
P_{I,\mathrm{shot}}(z) = \int d M_\mathrm{h} \frac{d n}{d M_\mathrm{h}} \left[ \frac{L(M_\mathrm{h},z)}{4\pi D^2_{L}} y(z) D^2_{A} \right]^2, 
\end{equation}
where $d n / d M_\mathrm{h}$ is the Sheth-Mo-Tormen halo mass function \citep{SMT2001} that our simulations are calibrated to and $D_L$ and $D_A$ are the luminosity and comoving angular diameter distances, respectively. The factor $y(z) = d \chi / d \nu$ maps the observed frequency $\nu$ into the comoving radial distance $\chi$. The total auto-power spectrum used for our ILI analysis is therefore the sum of the clustering and shot-noise components. 

\begin{table*}
\centering
\caption{Survey and instrument specifications of the $[\ion{C}{II}]$ and $[\ion{O}{III}]$ LIM surveys considered in this work. Reference values of $\Omega_\mathrm{beam}$, $V_\mathrm{vox}$, and $P_\mathrm{N}$ are given at $z=7$ (7.5 for ground-based $[\ion{O}{III}]$).}
\vspace{0.1cm}
\label{tb:sens}
\resizebox{\textwidth}{!}{%
\begin{tabular}{cccccccccccc}
\toprule
\textbf{Target} & \textbf{Redshifts} & \textbf{Bandpass} & $D_\mathrm{ap}$ & $\Omega_\mathrm{survey}$ & $t_\mathrm{survey}$ & $N_\mathrm{pix}$ & $R$ & $\Omega_\mathrm{beam}$ & $V_\mathrm{vox}$ & $\sigma_\mathrm{N}$ & $P_\mathrm{N}$ \\
 & & (GHz) & (m) & (deg$^2$) & (hr) & & & (arcmin$^2$) & (Mpc$^3$) & (Jy\,sr$^{-1}$\,s$^{1/2}$) & ((Jy\,sr$^{-1}$)$^2$\,Mpc$^{3}$) \\
\midrule
\multicolumn{12}{c}{\textit{Ground-based, FYST-like survey}} \\
\midrule
$[\ion{C}{II}]$ & 6, 7, 8, 9 & 190--270 & 10 & 4 & 4000 & 1000 & 100 & 0.3 & 67  & $2\times10^6$ & $3\times10^9$ \\
$[\ion{O}{III}]$ & 7.5, 8.5 & 360, 400 & 10 & 4 & 4000 & 1000 & 100 & 0.1 & 23  & $4\times10^6$ & $1.2\times10^{10}$ \\
\midrule
\multicolumn{12}{c}{\textit{Space-based, Padmanabhan+22-like survey}} \\
\midrule
$[\ion{C}{II}]$ & 6, 7, 8, 9 & 190--270 & 3 & 4 & 1000 & 100 & 100 & 3   & 750 & $2\times10^5$ & $1.2\times10^9$ \\
$[\ion{O}{III}]$ & 6, 7, 8, 9 & 340--490 & 3 & 4 & 1000 & 100 & 100 & 1   & 230 & $2\times10^5$ & $1.2\times10^9$ \\
\bottomrule
\end{tabular}%
}
\end{table*}

The simulated [$\ion{C}{II}$] and [$\ion{O}{III}$] intensity fluctuations, which roughly trace the underlying matter density distribution, are visualized for an example model in the bottom panels of figure~\ref{fig:rendering}. The corresponding auto-power spectra are shown in the top left panel, which includes the (analytically derived) shot-noise contribution responsible for the flattening at small scales. 

\subsubsection{Cross-power spectra with the galaxy distribution}

The cross-correlation between the line intensity field and the galaxy distribution is another useful signal to consider given its robustness to foregrounds like line interlopers and other observational systematics \cite{Chung2019,VisbalMcQuinn2023}. It enables clean measurements of both the (bias-weighted) line mean intensity on large scales and the average line luminosity of the galaxies being cross-correlated on small scales. To simulate this cross-correlation signal, we inspect halos identified through filtering the linear density field with a range of smoothing scales (i.e., the excursion set formalism; \cite{MesingerFurlanetto2007}) and select those brighter than a limiting magnitude, $m_\mathrm{AB,lim}$, converted from their SFR predicted by our galaxy model. The selected halos constitute our galaxy sample and their distribution is cross-correlated with the line intensity field. We note that on shot-noise-dominated scales, we take the same hybrid approach to evaluate the cross-shot-noise power spectrum analytically as
\begin{equation}
P_{I\times\mathrm{gal},\mathrm{shot}}(z) = \frac{\bar{I}_\mathrm{gal}(z)}{n_\mathrm{gal}(z)} \propto \bar{L}_\mathrm{gal}(z),
\end{equation}
where $\bar{I}_\mathrm{gal}$ ($\bar{L}_\mathrm{gal}$) is the mean line intensity (luminosity) of the galaxies involved in the cross-correlation. Specifically, in this work we consider a moderate depth survey of the Roman Space Telescope for which the 5$\sigma$ point source limiting magnitude is $m_\mathrm{AB,lim} = 28.2$ \citep{Yung2023}. 

We mark the locations of these Roman-detectable galaxies in figure~\ref{fig:rendering} for the example model and show power spectra of their auto-correlation and cross-correlation with the two respective lines. Both the clustering-dominated ($k \ll 1\,\mathrm{Mpc^{-1}}$) and shot-noise-dominated ($k \gg 1\,\mathrm{Mpc^{-1}}$) scales contain useful information about the galaxy formation parameters of interest. The top middle and top right panels show the line intensity--galaxy cross-power spectra and the auto-power spectrum of galaxies, respectively. 

\subsection{Observational effects} \label{sec:obs_effects}

To assess the impact of measurement uncertainties on our SBI analysis, we take into account two major observational effects on our power spectral analysis: the instrument noise and the finite survey resolution. Whereas the former introduces a scale-independent noise power spectrum, the latter attenuates the power spectrum signal in both spatial and spectral directions. Following \cite{Lidz2011}, we can express the variance of the observed auto-power spectrum for a single Fourier mode as
\begin{equation}
\mathrm{var}\left[\Delta^2_{I}(k, \mu)\right] = \left[ \Delta^2_{I}(k) + \Delta^2_\mathrm{N}(k) / W^{2}(k, \mu) \right]^2. 
\end{equation}
The instrument noise $\Delta^2_\mathrm{N}(k) = k^3 P_\mathrm{N} / 2\pi^2$, where
\begin{equation}
P_\mathrm{N} = V_\mathrm{vox} {\sigma}_\mathrm{N}^2/t_\mathrm{obs}  
\end{equation}
and $t_\mathrm{obs} = N_\mathrm{pix}(\Omega_\mathrm{beam}/\Omega_\mathrm{survey})t_\mathrm{survey}$ is the observing time per spatial pixel. The window function accounts for the finite spatial and spectral resolution, namely
\begin{equation}
W^2(k, \mu) = e^{-k^2 \sigma^2_{\perp} - k^2(\sigma^2_{\parallel} - \sigma^2_{\perp})\mu^2}, 
\end{equation}
with $\mu = \cos \theta$ for angle $\theta$ between the $k$ vector and the line of sight, $\sigma_{\parallel} = k^{-1}_\mathrm{\parallel,max} = d \chi/d \nu \delta \nu$, and $\sigma_{\perp} = k^{-1}_\mathrm{\perp,max} = \chi \Omega^{1/2}_\mathrm{beam}$ for spectral resolution $\delta \nu$, comoving radial distance $\chi$, and beam size $\Omega_\mathrm{beam}$. Taking into account the number of (independent) Fourier modes sampled $N_\mathrm{mode}(k) = 4\pi k^2 \Delta k V_\mathrm{survey}/(2\pi)^3/2 = k^2 \Delta k V_\mathrm{survey}/4\pi^2$ and with inverse-variance weighting, we have
\begin{equation}
\frac{1}{\mathrm{var}\left[ \Delta^2_{I}(k) \right]} = \int_{0}^{1} \frac{k^2 \Delta k V_\mathrm{survey}/4\pi^2}{\mathrm{var}\left[\Delta^2_{I}(k, \mu)\right]} d \mu. 
\label{eq:mu_weighting}
\end{equation}
Following the treatment in \cite{Zhao2022PS}, effects of the instrument noise can be taken into account by adding a Gaussian random offset with zero mean and variance, $\mathrm{var}[\Delta_{I}^2(k)]$ to the simulated power spectra. While in practice it will be more realistic and direct to incorporate these effects into simulated 3D line intensity data cubes, generating such fully survey-like synthetic observations requires detailed modeling of survey and instrument specifics, which is beyond the defined scope of this work to demonstrate the inference of galaxy formation physics from power spectrum constraints. We therefore leave to future work detailed simulations of LIM data in the observing frame, along with the exploration of field-level ILI. We should also emphasize that even under the assumption of Gaussian measurement errors on power spectra, ILI remains valuable because the forward simulation introduces strong nonlinearities in the mapping between galaxy formation parameters and power spectrum observables. Training on simulations assuming Gaussian power spectrum uncertainties thus enables the ILI framework to capture the full posterior structure while bypassing the need for an explicit likelihood that is difficult to evaluate in closed form.

For the line intensity--galaxy cross-power spectrum, the variance per mode can be similarly written as \cite{Lidz2009}
\begin{equation}
\mathrm{var}[\Delta^2_{\times}(k, \mu)] = \frac{\left[\Delta^2_{\times}(k) \right]^2 + \sqrt{\mathrm{var}\left[\Delta^2_{I}(k, \mu)\right] \mathrm{var}\left[\Delta^2_\mathrm{g}(k, \mu)\right]}}{2},
\end{equation}
where
\begin{equation}
\mathrm{var}\left[\Delta^2_\mathrm{g}(k, \mu)\right] = \left[\Delta^2_\mathrm{g}(k) + \Delta^2_\mathrm{g,P}(k)/W_\mathrm{g}^2(k,\mu)\right]^2
\end{equation}
and
\begin{equation}
W^2_\mathrm{g}(k,\mu) = e^{-k^2\sigma^2_\mathrm{\parallel,g}\mu^2} = e^{-k^2 [c \sigma_z / H(z)]^2 \mu^2}. 
\end{equation}
After inverse-variance weighting of $\mu$, the average variance, $\mathrm{var}\left[\Delta^2_\mathrm{\times}(k)\right]$, is related to the per mode variance, $\mathrm{var}\left[\Delta^2_\mathrm{\times}(k, \mu)\right]$, again by Equation~(\ref{eq:mu_weighting}). For the redshift uncertainty $\sigma_z$, we adopt $\sigma_z \approx 0.1 (1+z)$ as a reasonable estimate for photometric redshift surveys of $z\gtrsim6$ LBGs by the Roman Space Telescope \citep[see e.g.,][]{LaPlante2022}. 

As summarized in table~\ref{tb:sens}, the two survey designs we consider are chosen to be comparable to current-generation, ground-based surveys like FYST/CCAT-prime, or a hypothetical space-based survey proposed by \cite{Padmanabhan2022}. In the former case, a larger aperture (yielding a smaller beam), more pixels, and longer integration time are possible, although atmospheric transmission windows limit [$\ion{O}{III}$] observations to narrow redshift ranges around $z\sim7.5$ and 8.5. In the latter case, both [$\ion{C}{II}$] and [$\ion{O}{III}$] can be mapped continuously over $6<z<9$ at a lower instrument noise level from space, albeit with a larger beam size, fewer pixels, and shorter observing time. It is therefore not guaranteed that space-based surveys will always outperform ground-based ones (see, for example, how ground-based experiments can achieve tighter [$\ion{C}{II}$]-based parameter constraints in figures~\ref{fig:posterior} and \ref{fig:fisher}). 

\begin{figure*}
 \centering
 \includegraphics[width=0.85\textwidth]{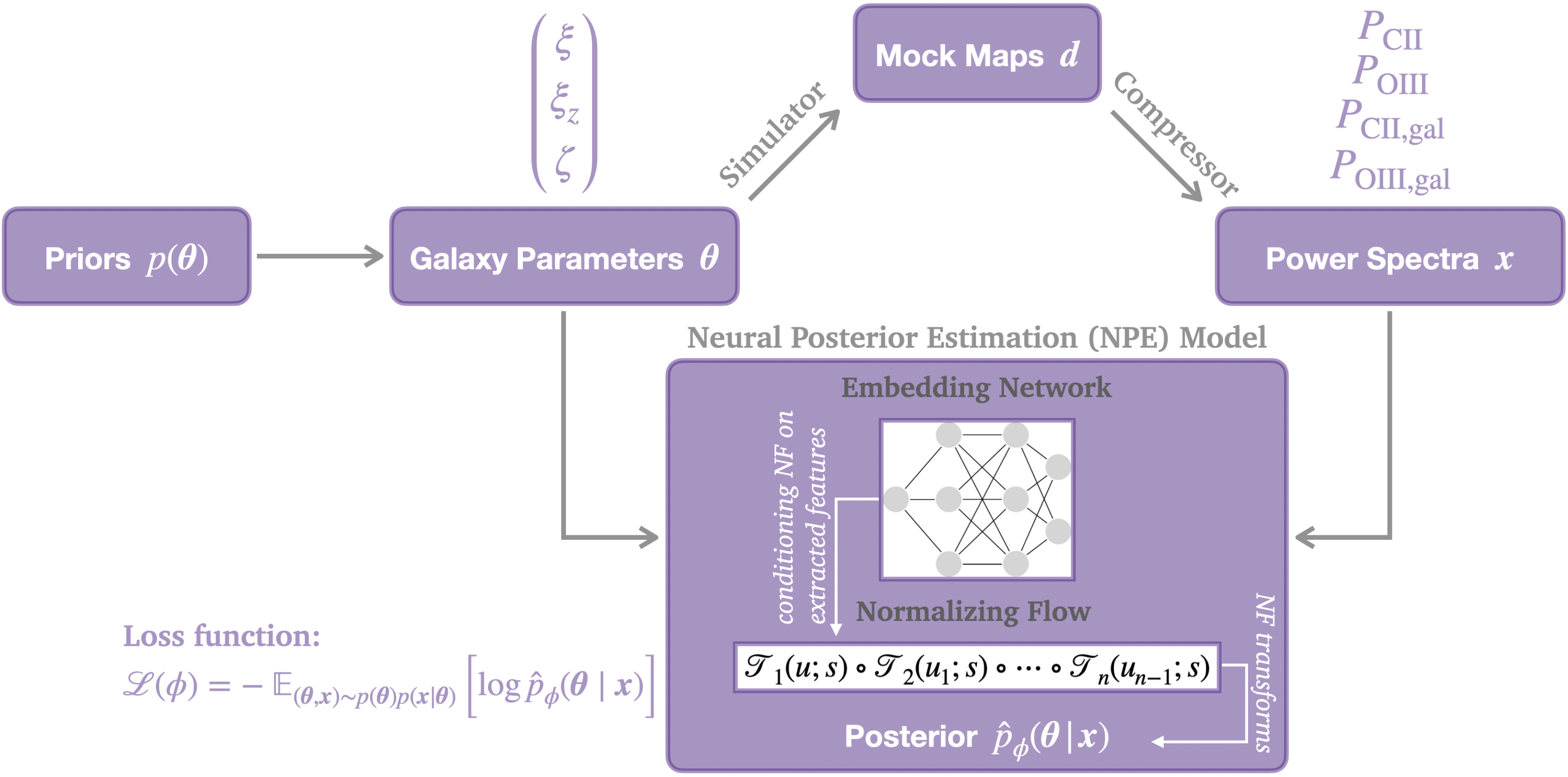}
 \caption{A schematic visualization of the workflow for the training process in the ILI framework presented. A neural posterior estimation (NPE) model is trained to learn the joint posterior of galaxy formation parameter vector $\boldsymbol{\theta} = (\xi, \xi_z, \zeta)$ from [$\ion{C}{II}$] and [$\ion{O}{III}$] auto-power spectra or their cross-power spectra with LBGs simulated by \texttt{LIMFAST}. The NPE model extracts informative features $(s)$ from power spectrum data using an embedding network (GRU; see appendix~\ref{sec:ml} for detail)  and use them to condition a normalizing flow (NF) that learns the target posterior by applying a sequence of invertible transforms $(\mathcal{T})$ to a simple base distribution $(u)$. Uniform priors are assumed for the parameters of interest. During inference, power spectra are supplied to the NPE model and the posterior is sampled without re-training.}
 \label{fig:workflow}
\end{figure*}


\section{The Implicit Likelihood Inference Framework} \label{sec:ili}

Bayesian inference lays the foundations for a statistical framework for parameter estimation in cosmological data analysis. Given an observed data set $\boldsymbol{x}$ and a vector of model parameters $\boldsymbol{\theta}$, Bayes' theorem relates the posterior distribution $p(\boldsymbol{\theta}|\boldsymbol{x})$ of interest to the product of the likelihood function $p(\boldsymbol{x}|\boldsymbol{\theta})$ and the prior distribution $p(\boldsymbol{\theta})$, namely
\begin{equation}
p(\boldsymbol{\theta}|\boldsymbol{x}) = \frac{p(\boldsymbol{x}|\boldsymbol{\theta}) p(\boldsymbol{\theta})}{p(\boldsymbol{x})},
\end{equation}
where the marginalized likelihood $p(\boldsymbol{x})$ normalizes the posterior. Traditional Bayesian inference methods rely on an explicit form of $p(\boldsymbol{x}|\boldsymbol{\theta})$, which, in some cases, can be analytically intractable or computationally expensive to evaluate, even if it is possible to forward simulate synthetic data. Such challenges are addressed by ILI, which bypasses the need for an explicit $p(\boldsymbol{x}|\boldsymbol{\theta})$ in a data-driven way by leveraging simulations to learn the joint distribution of simulated data-parameter pairs. This provides a more flexible and scalable approach to Bayesian inference less constrained by model-specific assumptions, and has therefore gained much attention in cosmological data analysis in recent years. Our LIMFAST simulations, which approximate structure formation and evolution using second-order Lagrangian perturbation theory and the extended Press--Schechter formalism, define an appropriate regime for the usage of ILI, where closed-form expressions for the observables are lacking and the large number of realizations required by methods such as on-the-fly MCMC would be computationally very demanding (without the aid of emulation) \citep{GreigMesinger2015}. By training on parameter--power spectrum data vector pairs simulated with LIMFAST, the ILI method learns to approximate the posterior, $p(\boldsymbol{\theta}|\boldsymbol{x})$, from the empirical joint distribution, implicitly exploiting the full shape and structure of the simulated power spectra and without requiring an explicit mapping from $\boldsymbol{\theta}$ to $\boldsymbol{x}$ as a forward emulator would.

In figure~\ref{fig:workflow}, we show the workflow of training the ILI framework that we use to infer the galaxy formation parameters of interest from mock [$\ion{C}{II}$] and [$\ion{O}{III}$] LIM data. To perform ILI, we train a Neural Posterior Estimation (NPE) model to learn the conditional distribution of the galaxy formation parameters given simulated data (compressed into auto- and cross-correlation power spectra) following the definitions in section~\ref{sec:models}. 
The NPE architecture first produces a compact embedding of the input data and then employs a normalizing flow \citep{PapamakariosMurray2016,Papamakarios2019} to approximate the complex, high-dimensional posterior by applying a sequence of invertible and differentiable transforms to a simple base distribution. Further details of the architecture and training of the NPE model in our ILI framework are elaborated in appendix~\ref{sec:ml}.  

To confirm the reliability of the NPE model in our ILI framework, we employ several diagnostics to validate its global performance in terms of the predictiveness and coverage of the learned posteriors, confirming that they are both informative and statistically well-calibrated across the full range of simulated realizations. Specifically, using our testing data, we compare true parameters to posterior predictions to visualize the constraining power. We use the probability--probability (P--P) plot to assess the calibration of posteriors by comparing the cumulative distribution function of true parameters against the predicted percentiles. In addition, we also leverage the Tests of Accuracy with Random Points (TARP; \cite{Lemos2023}) method to perform a flexible, sampling-based coverage test of the generative posterior estimators. Without requiring explicit evaluation of the posterior, TARP calculates the distance between random samples drawn from the multi-dimensional parameter space and the true values to test the accuracy of posterior estimators and has been shown to provide accurate coverage estimates given a sufficient number of samples. Together, these tests offer a robust validation of the ILI framework and its ability to generalize across our interested parameter space. 


\section{Results} \label{sec:results}

\subsection{Summary of the simulated data}

For our ILI analysis, we use the Latin Hypercube Sampling (LHS) to generate 1000 randomly distributed samples of $(\xi, \xi_z, \zeta)$ from the parameter space $0 \leq \xi \leq 1$ (for the mass loading factor's mass dependence), $-1 \leq \xi_z \leq 1$ (for the mass loading factor's redshift dependence), and $1 \leq \zeta \leq 2$ (for the slope of the $\dot{\Sigma}_{\star}$--$\Sigma_\mathrm{g}$ relation), assuming bounded uniform priors. For each sample drawn from the three-dimensional parameter space, we run one \texttt{LIMFAST} simulation from $z=20$ to $z=5$ to create the co-eval boxes of [$\ion{C}{II}$] and [$\ion{O}{III}$] intensities (in the density field mode) and the spatial distribution of Roman LBGs (in the halo field mode) using same initial conditions. We then calculate the auto- and cross-correlation power spectrum signals from the simulated boxes at the relevant redshifts (e.g., $z=7.5$ and 8.5 for ground-based [$\ion{O}{III}$] observations) in 10 $k$ bins with a fixed bin width of $\Delta \log k = 0.15$. The simulated dataset after compression is partitioned into training, validation, and test subsets in an $8:1:1$ ratio, respectively. We create 10 random realizations of the noise-injected power spectra for each sample assuming Gaussian uncertainties (see section~\ref{sec:obs_effects}). In each realization, we neglect the covariance between different redshifts, as well as between different tracers when both lines are involved. In other words, when injecting noise into the simulated power spectra, we assume that the noise realizations for different lines are independent: although the two lines trace the same large-scale structure at a given redshift, we expect the covariance between their measured power spectra to be relatively weak unless the measurements are strongly cosmic variance dominated, which motivates this simplified treatment. A more realistic treatment that includes the full multi-line covariance would require more involved astrophysical modeling of the line--line correlation (see e.g., \citep{Yang2022}) and a detailed consideration of their separation (from each other and from other foreground components). The input data matrix therefore has a dimension of (10000, 10, $N_{z}N_{I}$), where $N_z$ ($N_I$) is the number of redshifts (lines) for which the power spectrum is measured. 

\begin{figure}[!ht]
 \centering
 \includegraphics[width=0.8\columnwidth]{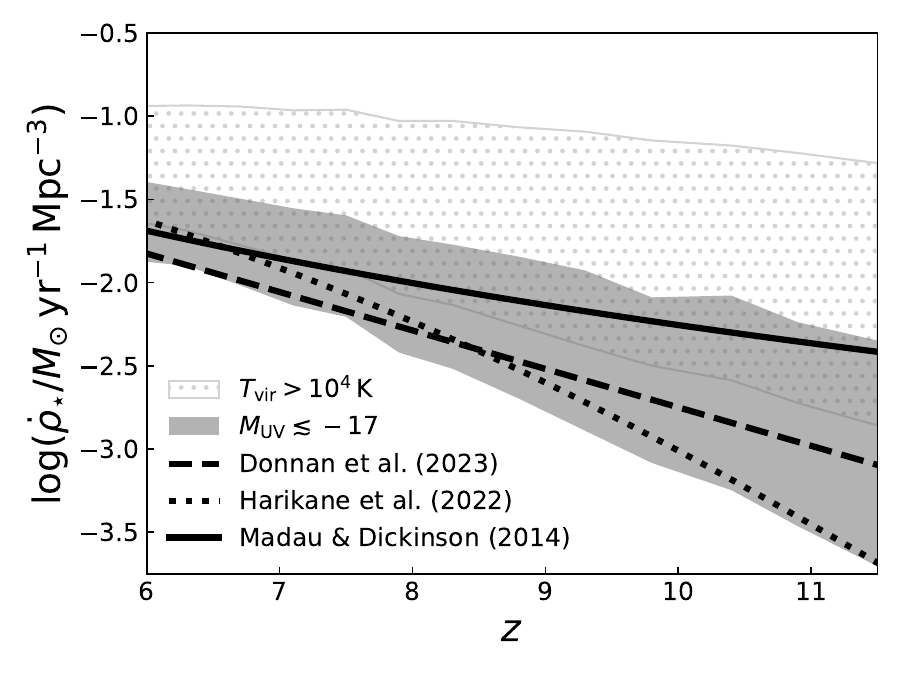}
 \caption{Comparison between the high-$z$ cosmic SFRDs sampled from our three-dimensional parameter space of interest (the 16--84th percentile) and the predictions from pre- and post-JWST empirical models \citep{MD2014,Harikane2022,Donnan2023}. The dark shaded and light hatched regions are integrated down to halos with virial temperature $T_\mathrm{vir}=10^4\,$K and absolute UV magnitude $M_\mathrm{UV} \simeq -17$, respectively, the latter of which corresponds to the detection limit assumed by the results from the literature. Note that, unlike the two post-JWST models that are constrained by measurements extending to $z>10$, \cite{MD2014} rely on extrapolation beyond $z>8$.}
 \label{fig:sfrd}
\end{figure}

\subsubsection{Cosmic SFRD}

The cosmic star formation rate density (SFRD) serves as a baseline sanity check for the range of galaxy formation model variations considered in our ILI analysis. In figure~\ref{fig:sfrd}, we show how the cosmic SFRDs simulated with our parameter samples compare against the predictions by either pre-JWST \citep{MD2014} or post-JWST \citep{Harikane2022,Donnan2023} empirical models. The total cosmic SFRD in our simulations as marked by the dotted hatched region is integrated down to halos with $T_\mathrm{vir}=10^4\,$K (roughly corresponding to the atomic cooling limit at the redshifts of interest). As a result, it lies above all the literature models, which show the SFRD extrapolated down to a limiting magnitude of $M_\mathrm{UV} \sim -17$ (which corresponds to $T_\mathrm{vir} \sim 3\times10^5\,$K) given the observed galaxy population. When the same integration limit as the observations is adopted, the predicted SFRDs align well with the literature models, as shown by the gray shaded band. These comparisons validate our baseline high-$z$ galaxy formation framework and the parameter ranges considered.  

\subsubsection{LIM observables}

\begin{figure*}[!ht]
 \centering
 \includegraphics[width=0.49\textwidth]{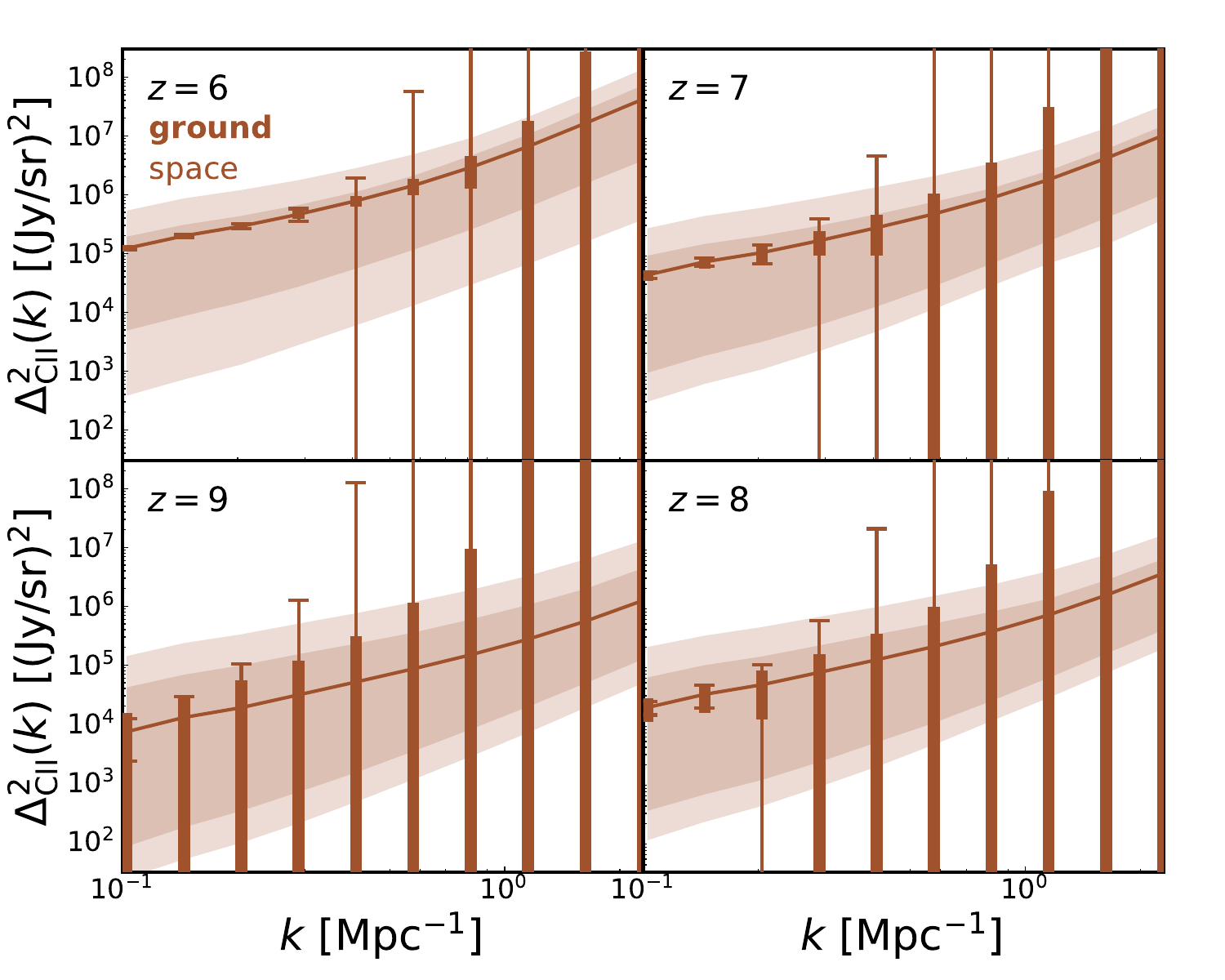}
 \includegraphics[width=0.49\textwidth]{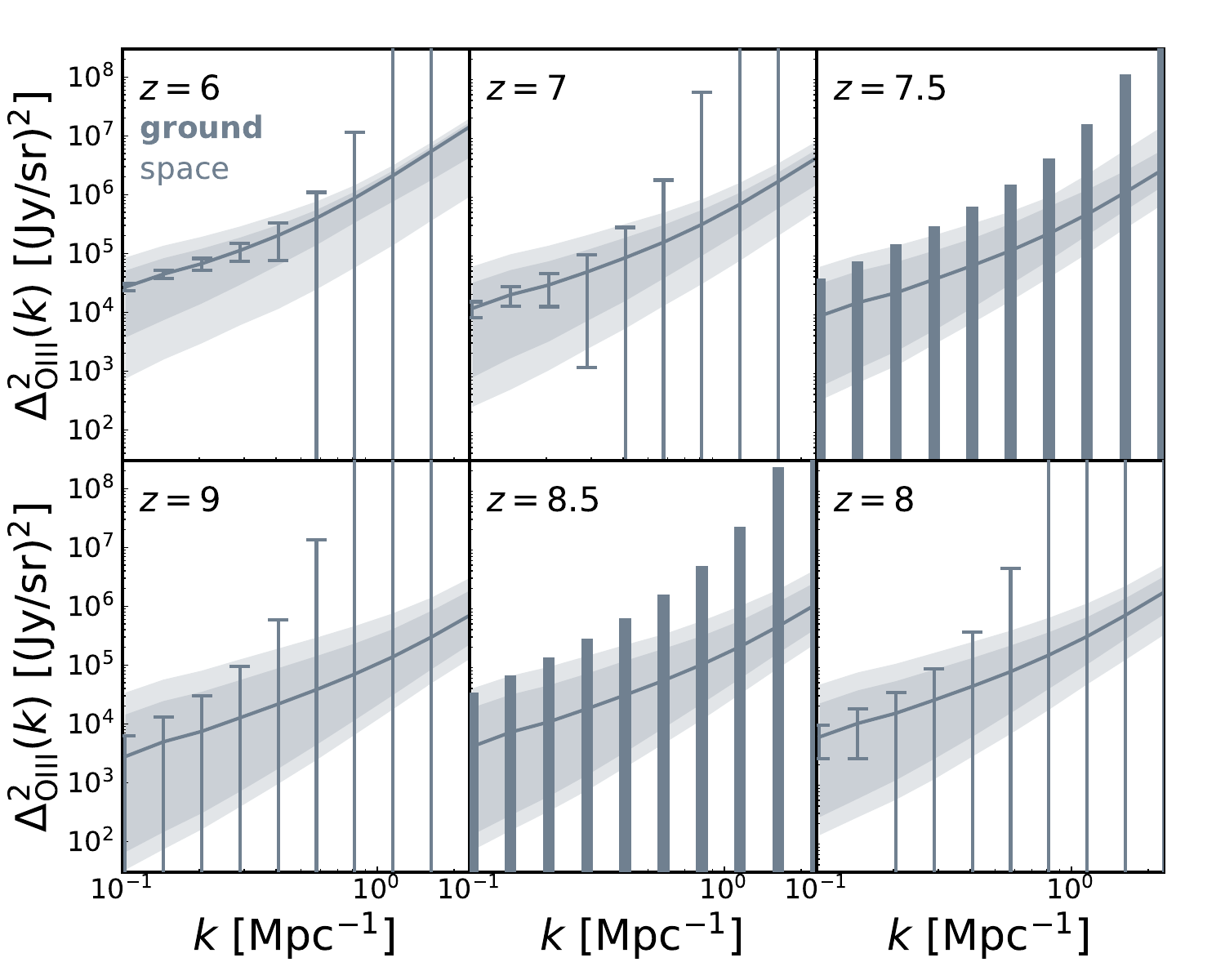}
 \caption{Samples of [$\ion{C}{II}$] (left) and [$\ion{O}{III}$] (right) auto-power spectra used for training our ILI framework. At each redshift, shaded bands in the background show the 16--84th and 5--95th percentiles of the samples drawn from the three-dimensional parameter space $0 \leq \xi \leq 1$ (for the mass loading factor's mass dependence), $-1 \leq \xi_z \leq 1$ (for the mass loading factor's redshift dependence), and $1 \leq \zeta \leq 2$ (for the slope of the $\dot{\Sigma}_{\star}$--$\Sigma_\mathrm{g}$ relation) of interest. Two sets of error bars are shown to illustrate the binned signal-to-noise levels predicted for the same example model shown in figure~\ref{fig:rendering} with $\xi \approx 1/3$, $\xi_z \approx 0$, and $\zeta \approx 1.4$, assuming the two experimental designs specified (``ground-based'' vs ``space-based'') in table~\ref{tb:sens}. The uncertainty blows up on small scales due to the finite spatial and spectral resolution of LIM surveys (section~\ref{sec:obs_effects}).}
 \label{fig:visualize_aps}
\end{figure*}

\begin{figure*}[!ht]
 \centering
 \includegraphics[width=0.49\textwidth]{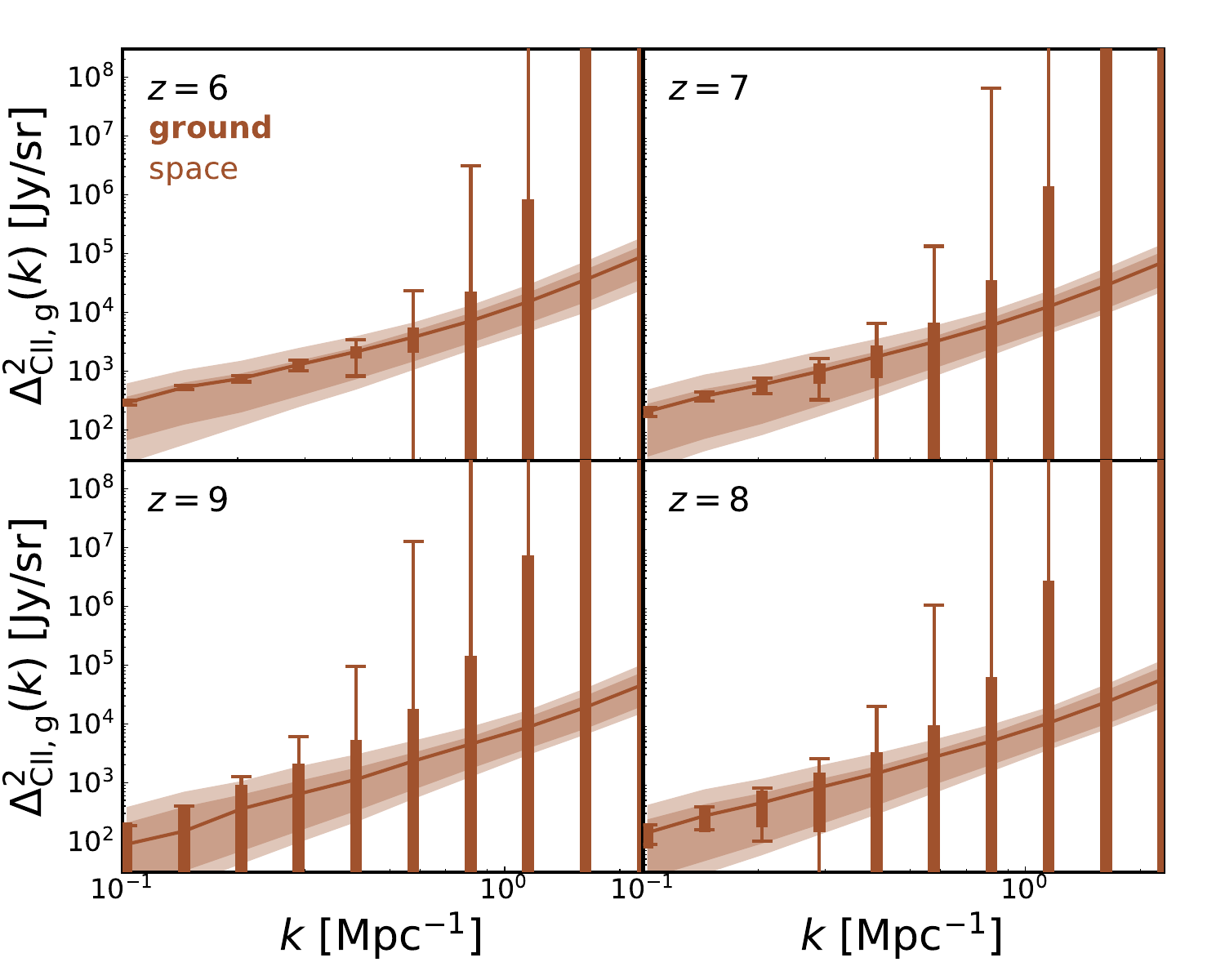}
 \includegraphics[width=0.49\textwidth]{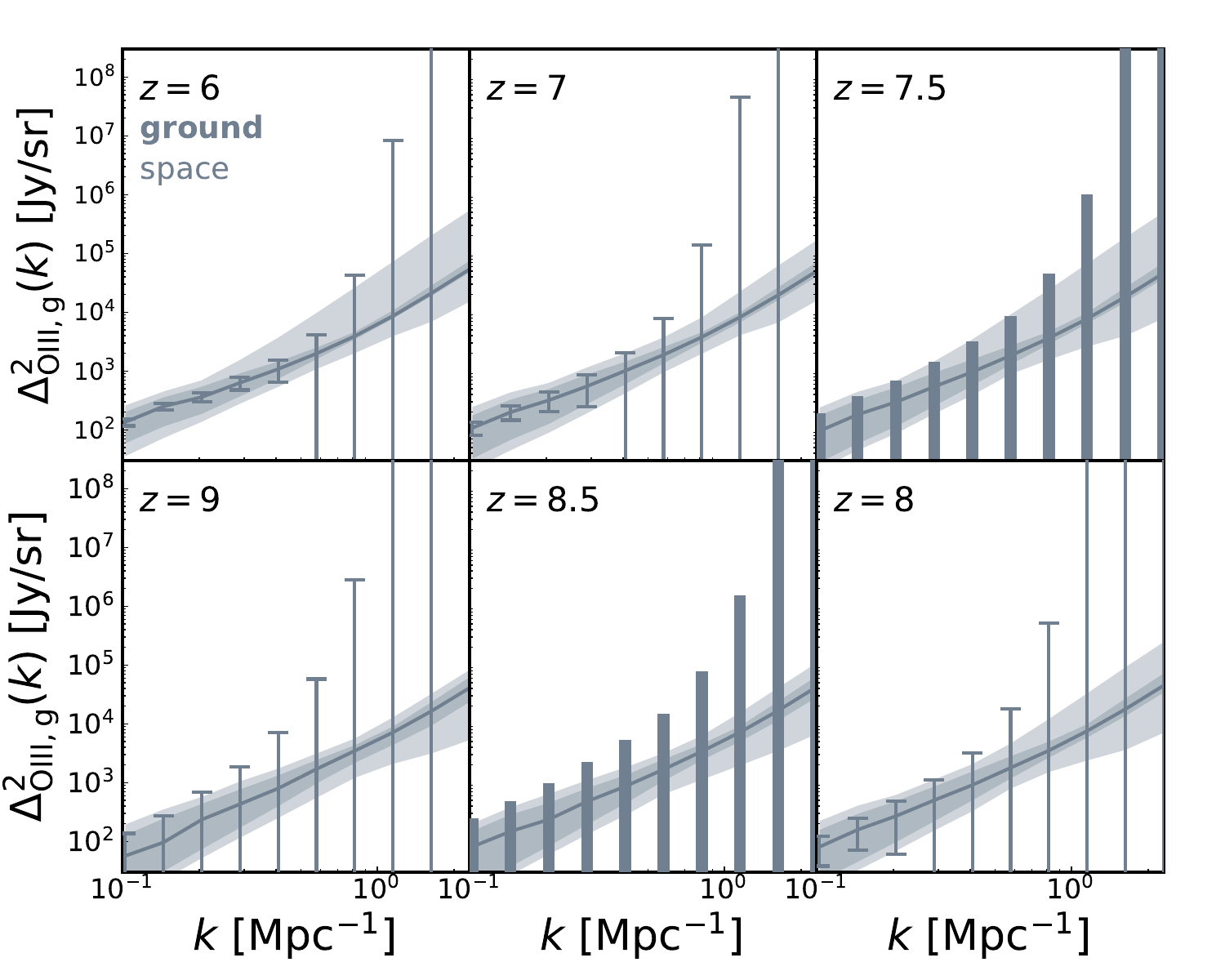}
 \caption{The same as figure~\ref{fig:visualize_aps}, but for line intensity--galaxy cross-power spectra.}
 \label{fig:visualize_cps}
\end{figure*}

We use \texttt{LIMFAST} to simulate four mock LIM datasets involving varying summary statistics (auto- vs cross-correlation) and survey specifications (ground- vs space-based experiment). Figures~\ref{fig:visualize_aps} and \ref{fig:visualize_cps} show auto-power spectra of [$\ion{C}{II}$] and [$\ion{O}{III}$] lines and their cross-power spectra with Roman LBGs, respectively. As illustrated by the shaded bands, the dynamic range associated with the sampled model variations is significantly narrower for the cross-correlation signals. This is because, compared with line intensities, the galaxy number statistics are less affected by the parameter variations. The predicted signal-to-noise (S/N) levels of power spectra in individual $k$ bins are indicated by the errorbars for an example model, in which case the auto- and cross-correlations are expected to be similarly detectable. We note, however, that in practice auto-power spectra are often more vulnerable to foreground contamination, including the confusion by interloping lines, which makes cross-power spectra the more robust observable for analysis although the information they probe is not identical. The two types of errorbars represent predictions for the current-generation ground-based and the next-generation space-based surveys, respectively. For the example model shown, the [$\ion{C}{II}$] power spectrum ([$\ion{C}{II}$]--galaxy cross-power spectrum) at $z=6$ can be measured at a total $\mathrm{S/N} = 20$ (17) and 26 (18) by the ground- and space-based experiments, respectively. [$\ion{O}{III}$] can only be measured by the space-based experiment, with a total $\mathrm{S/N} = 11$ (13) at $z=6$ in auto-correlation (cross-correlation). Note that these numbers are provided only as rough estimates of the detectability for this example model---one among many possibilities---and the actual constraints on these power spectra may be significantly weaker due to unaccounted practical challenges, such as foreground subtraction and component separation (see section~\ref{sec:discussion}). Finally, although not shown explicitly, we have verified that the predicted [$\ion{C}{II}$] and [$\ion{O}{III}$] power spectra are generally consistent with previous empirical models calibrated against observed scaling relations \cite{Padmanabhan2022}. 

\subsection{Joint parameter constraints from ILI}

\begin{figure*}[!ht]
 \centering
 \includegraphics[width=0.49\textwidth]{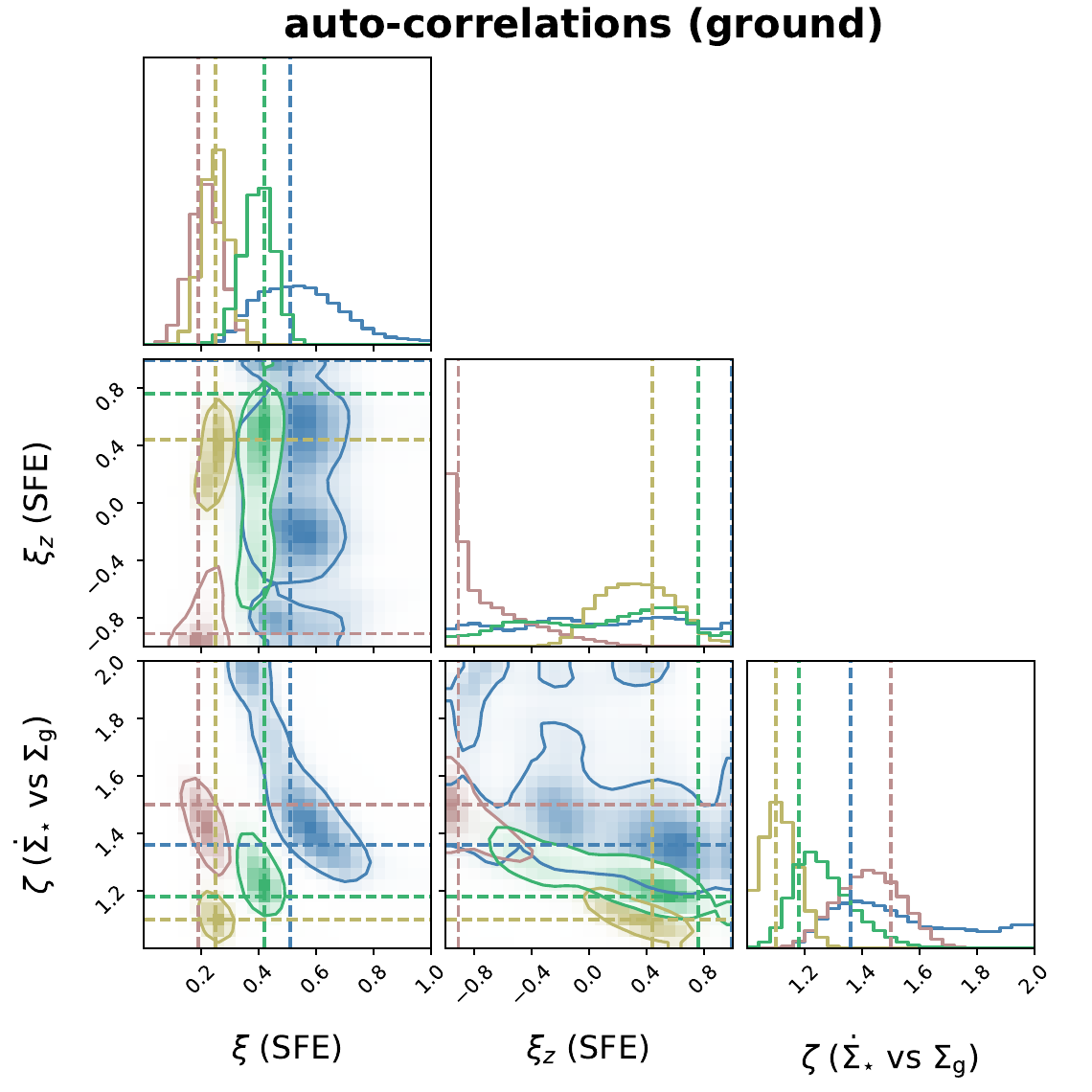}
 \includegraphics[width=0.49\textwidth]{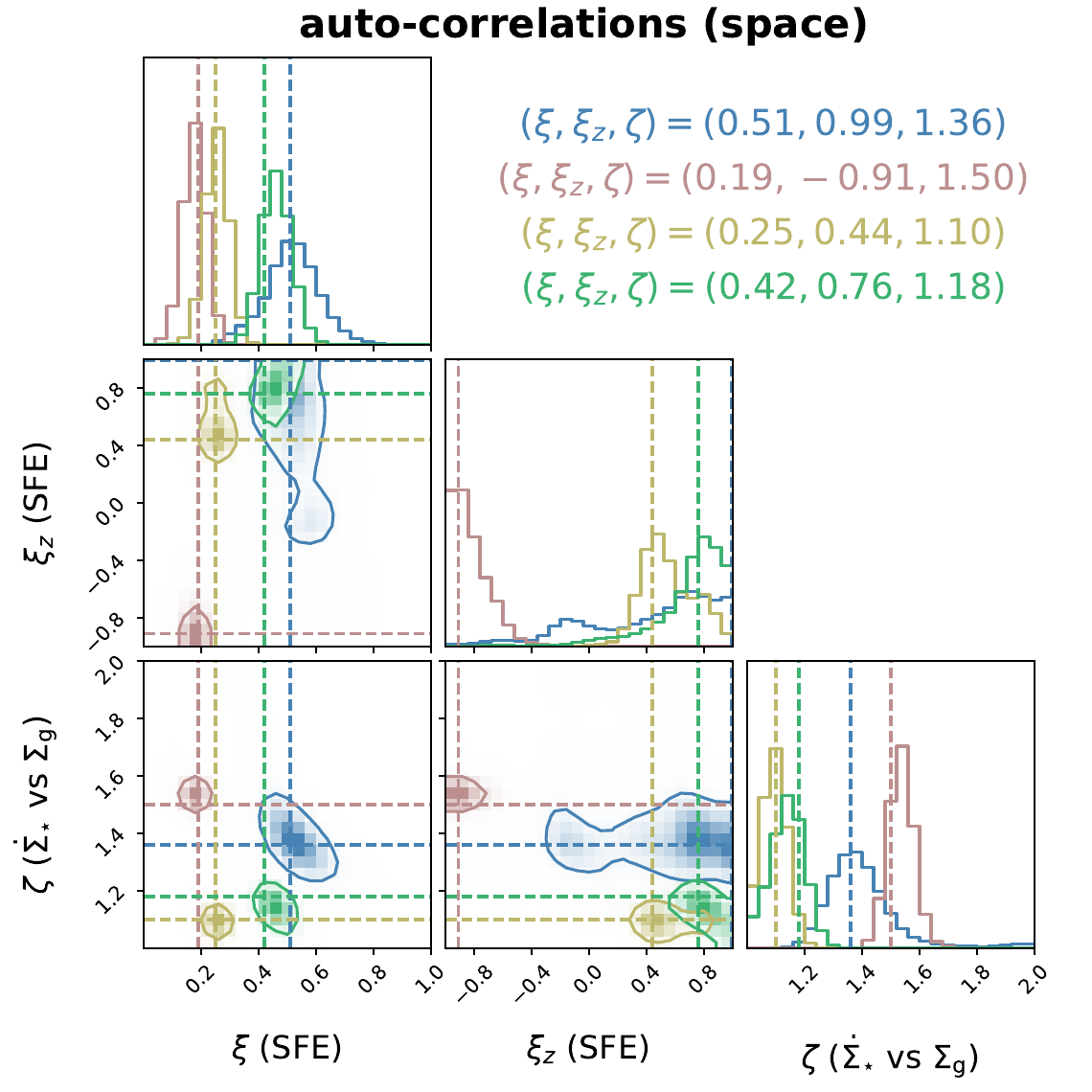}
 \includegraphics[width=0.49\textwidth]{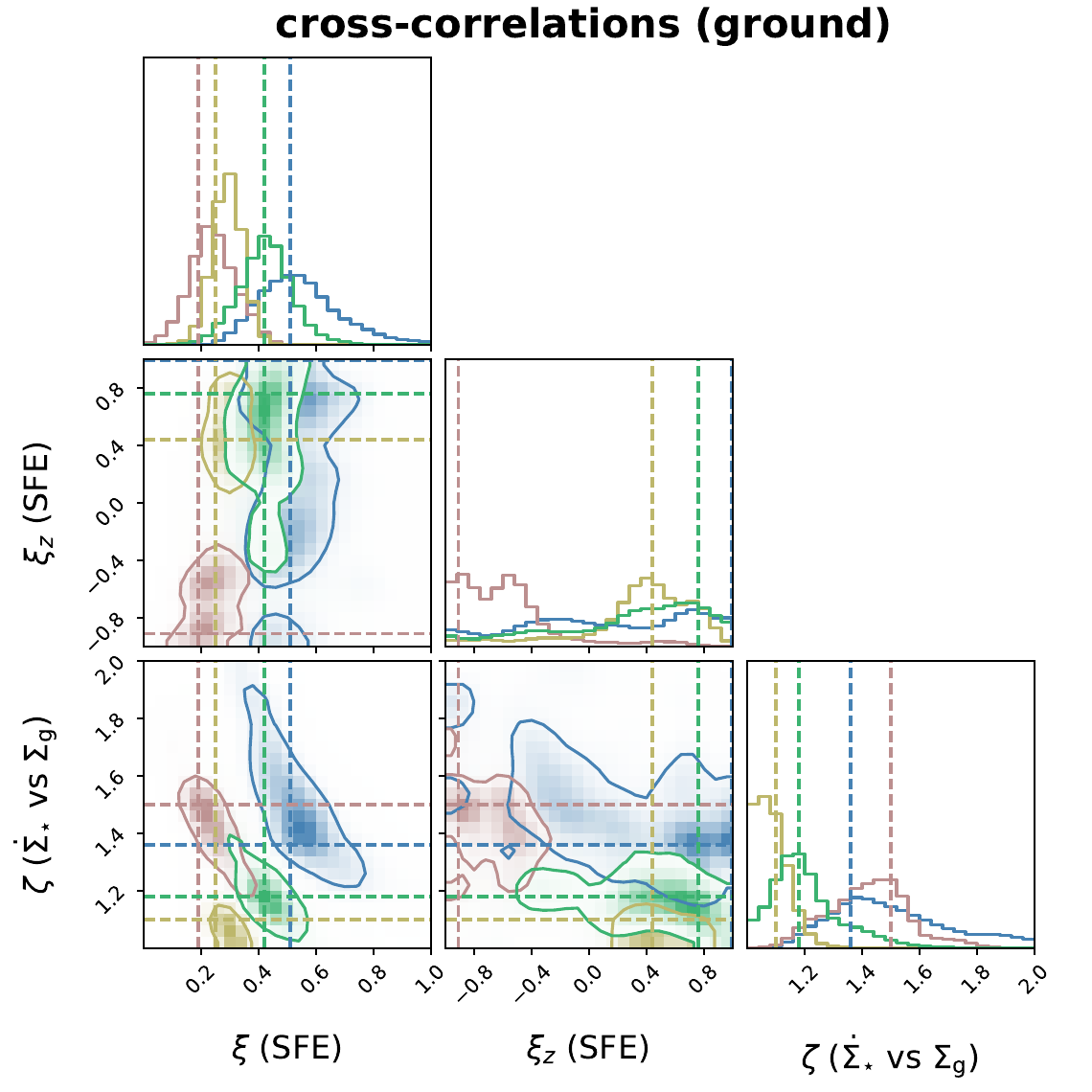}
 \includegraphics[width=0.49\textwidth]{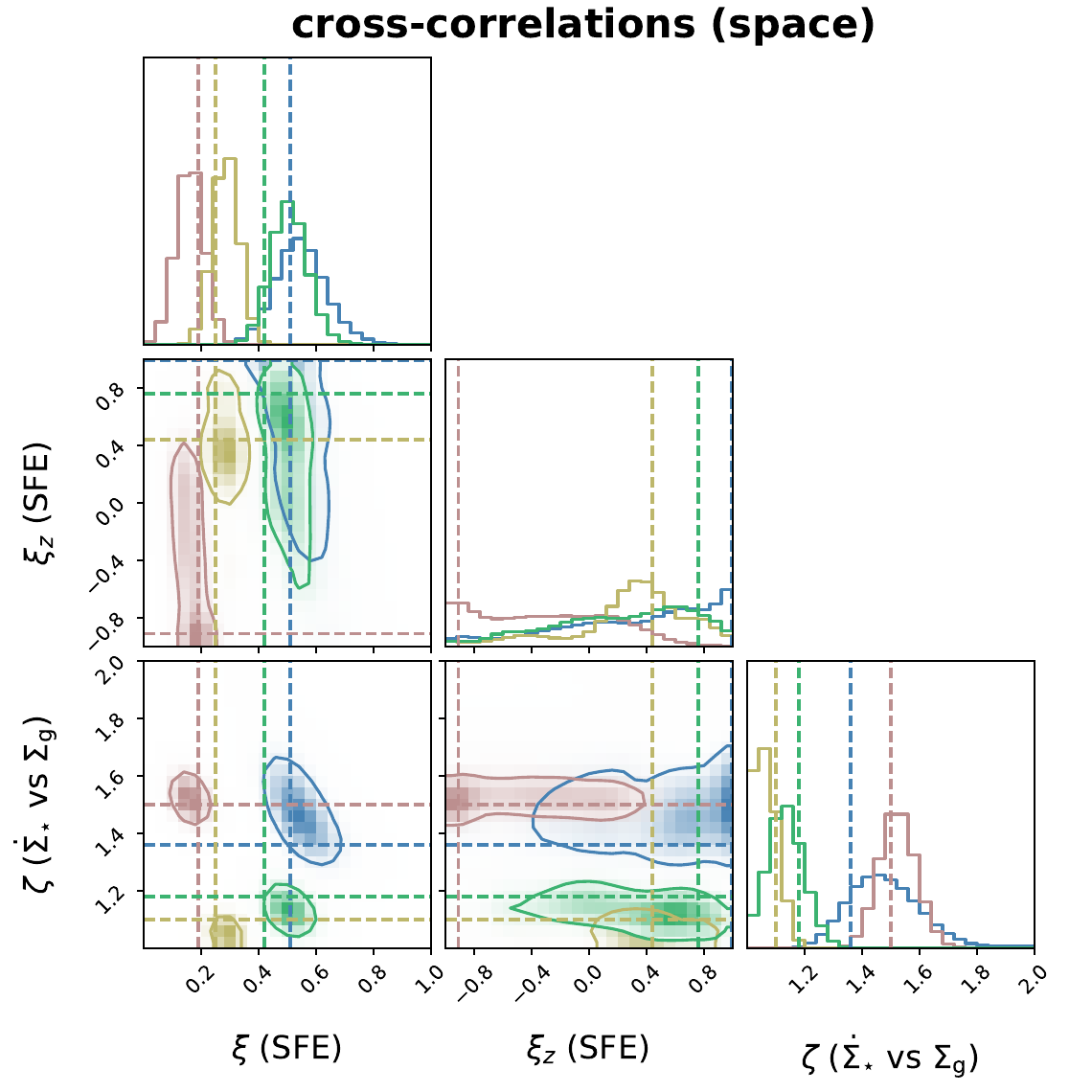}
 \caption{Posteriors of the three galaxy formation parameters of interest inferred through our ILI framework using both [$\ion{C}{II}$] and [$\ion{O}{III}$] lines but for different combinations of summary statistics and survey specifications. Four example models spanning the parameter space are shown in different colors, and their true values are indicated by the crosshairs.}
 \label{fig:posterior}
\end{figure*}

\begin{figure*}[!ht]
 \centering
 \includegraphics[width=0.75\textwidth]{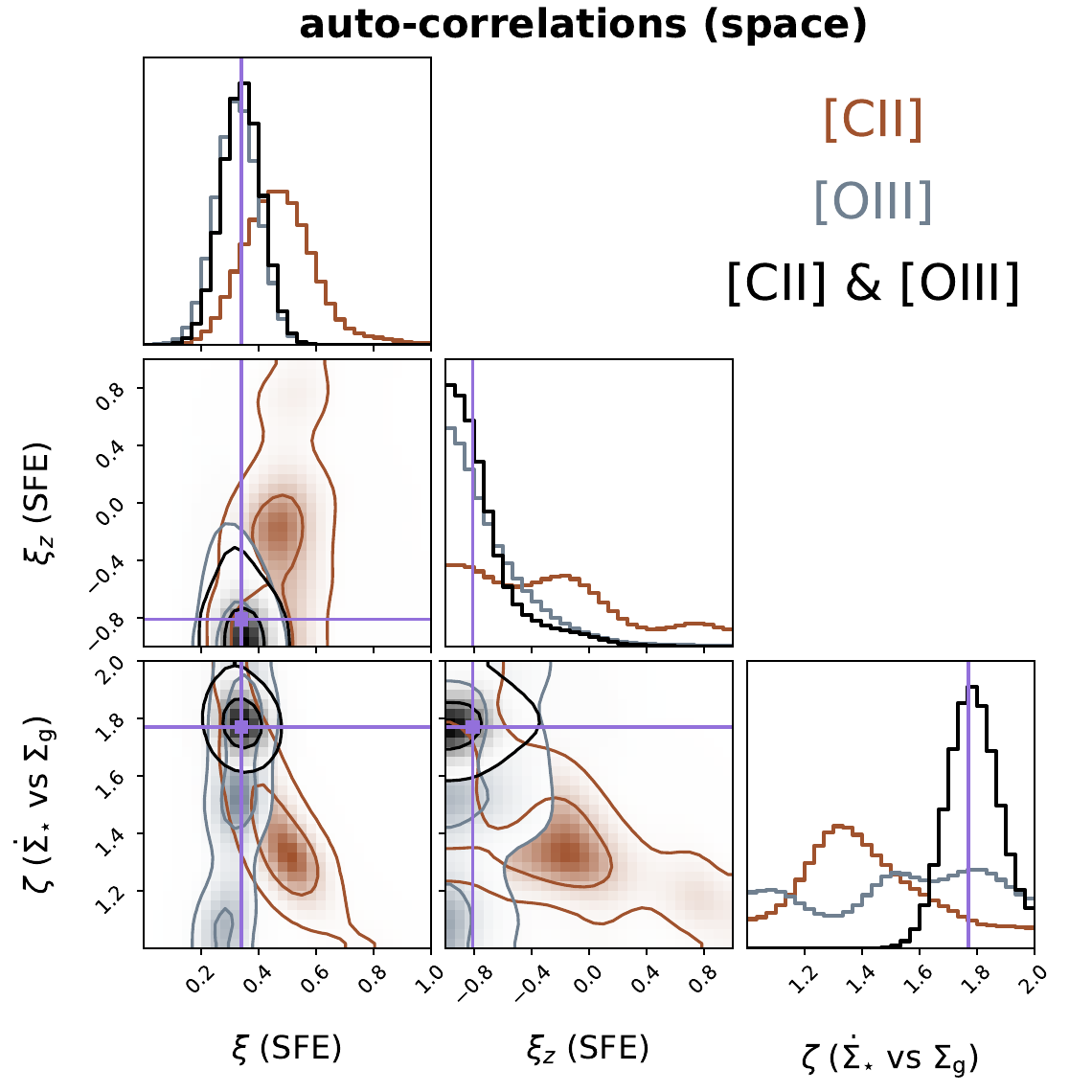}
 \caption{Similar to figure~\ref{fig:posterior} but showing here the comparison of posteriors inferred from single vs. both line tracers for one example model.}
 \label{fig:posterior_cii_oiii_both}
\end{figure*}

Figure~\ref{fig:posterior} showcases the results from our ILI analysis using both [$\ion{C}{II}$] and [$\ion{O}{III}$] lines, obtained by evaluating the learned posterior on the testing data that were excluded from training and validation. We define four scenarios based on the choice of summary statistics---either line auto-power spectra or line–galaxy cross-power spectra---and the assumed survey specifications, corresponding to the ground- or space-based experiment as specified in table~\ref{tb:sens}. For each scenario, we examine and compare the joint posteriors evaluated for four different sets of the parameters spanning the parameter space. We remind that both the training of the NPE model (including the optimization of hyperparameters) and the inference are done independently for each scenario (i.e., auto vs. cross and ground vs. space). Notably, in all four scenarios, the inferred posteriors recover the true values without significant bias, although the achieved constraints vary across the scenarios due to the different levels of uncertainties in the summary statistics. We also emphasize that the NPE model is trained with the same amount of 8000 samples in all these scenarios. However, as shown by figure~\ref{fig:scaling} in appendix~\ref{sec:ml}, training becomes more challenging and requires larger sample sizes for noisier measurements (e.g., ground-based cross-correlations). As a result, for the same training sample size, configurations or parameter sets with higher uncertainties are less well constrained and their inferred posteriors can show spurious shapes and features due to noisy data and sample-limited training. 

In addition to results inferred from both lines, we also show in figure~\ref{fig:posterior_cii_oiii_both} a comparison of posteriors obtained from single vs. both lines, which reveals several noteworthy insights. First, with [$\ion{C}{II}$] LIM measurements alone, tight constraints on both $\xi$ and $\zeta$ are prevented by their strongly degenerate effects on the [$\ion{C}{II}$] luminosity--halo mass relation. Next, incorporating [$\ion{O}{III}$] measurements enabled by the space-based experiment significantly tightens the constraints on both $\xi$ and $\zeta$, thanks to the distinct degeneracy directions of [$\ion{C}{II}$] and [$\ion{O}{III}$] given the different ways they depend on halo properties. As shown by the comparison between posteriors in black and brown/gray, even though constraints from each individual line remain loose due to significant degeneracies, the combined constraints become remarkably tight once these degeneracies are lifted. For instance, the correlation coefficient between $\xi$ and $\zeta$ changes from $\sim -0.9$ to $\sim -0.6$ when [$\ion{O}{III}$] is combined with [$\ion{C}{II}$] data. A joint analysis of both lines therefore proves essential for placing tight constraints on $\zeta$, which remains otherwise weakly constrained. We note that similar insights can be drawn from either auto- or cross-correlation analyses, although for a fixed, limited sample size, both the performance of training and the resulting posteriors may be affected by the level of uncertainties in the data. 

For comparison, and as a sanity check, we also derive the parameter constraints in appendix~\ref{sec:fisher} using the Fisher matrix, for which we generate simulations by offsetting the three parameters of interest one at a time from their reference values. As illustrated by figure~\ref{fig:fisher}, the Fisher matrix predicts qualitatively similar constraints and parameter degeneracies. This corroborates our main finding that combining [$\ion{C}{II}$] and [$\ion{O}{III}$] LIM measurements is crucial for breaking the degeneracy between the star formation efficiency ($\xi$) and the $\dot{\Sigma}_{\star}$--$\Sigma_\mathrm{g}$ relation ($\zeta$), thereby enabling robust constraints on the latter. We caution, however, that these results only include instrumental noise given the adopted survey specifications, and do not account for contamination from continuum emission or interloping line foregrounds (including the confusion between [$\ion{C}{II}$] and [$\ion{O}{III}$] lines themselves). Addressing these challenges is left for future work, but we emphasize that the need of foreground mitigation and component separation could substantially weaken the parameter constraints that can be achieved.

\subsection{Validation of the inferred posteriors} \label{sec:results:validation}

Using the metrics introduced in section~\ref{sec:ili}, we validate the inferred posteriors, quantifying their predictiveness and coverage across the parameter space of interest. For each individual case in a given scenario shown in figure~\ref{fig:posterior}, we compare true vs. predicted parameter values and assess the calibration of the inferred posteriors using the P--P plot and the sampling-based TARP method \citep{Lemos2023}. Specifically, the P--P plot displays the empirical cumulative distribution function (CDF) of the inferred posterior from probability integral transform (PIT) versus the nominal probability level. The comparison between the curve and the 1:1 diagonal informs whether the inferred posterior is biased (systematically offset high or low), overconfident (bulging away from the diagonal center), or underconfident (bowing toward the diagonal center). TARP, on the other hand, draws an analogous validation curve vs. diagonal comparison, but instead compares the expected coverage (estimated based on the distance between the true values and randomly sampled reference points) of the posterior samples against the credibility level. It has been proven to provide a necessary and sufficient multivariate test of full-posterior accuracy. Whereas proper PIT-based coverage tests can require exponentially more samples as the dimensionality of the parameter space grows, TARP is shown to accurately approximate and assess the true posterior coverage at much lower computational cost. As in the case of the P--P plot, a TARP curve close to the 1:1 diagonal suggests that the posterior is well-calibrated (i.e., neither overconstrained nor underconstrained).

\begin{figure*}[!ht]
 \centering
 \includegraphics[width=\textwidth]{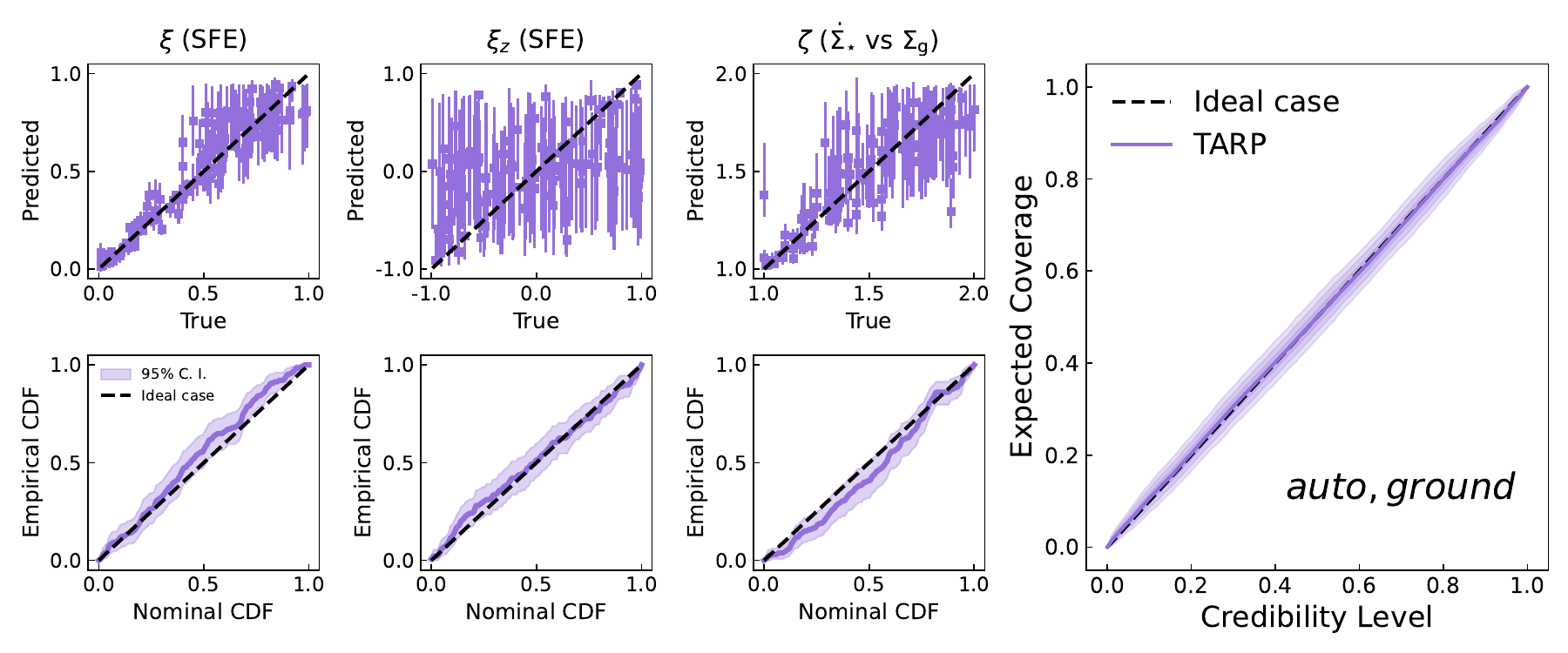}
 \includegraphics[width=\textwidth]{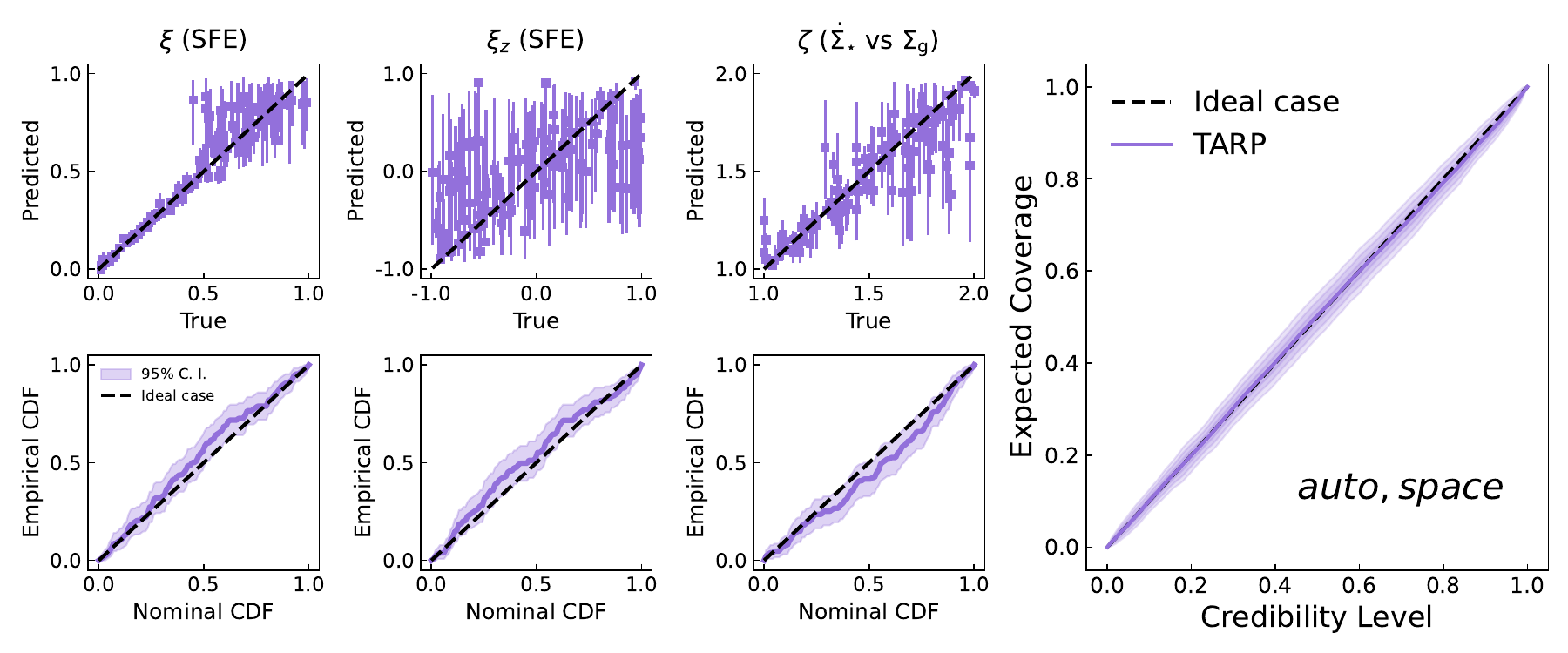}
 \caption{Validation of the predictiveness and calibration of the posteriors inferred by the ILI framework for cases where [$\ion{C}{II}$] and [$\ion{O}{III}$] auto-power spectra are measured by the current-generation, ground-based (top) and the hypothesized space-based (bottom) experiments, respectively. Three different measures are shown, including the comparison between true and predicted parameter values, the P-P plot comparing the true and predicted percentiles, and the TARP curve (see text for details).}
 \label{fig:validation}
\end{figure*}

In figure~\ref{fig:validation}, we show examples of these validation tests for two cases shown in figure~\ref{fig:posterior}, where the posterior is inferred from auto-power spectra of both [$\ion{C}{II}$] and [$\ion{O}{III}$] lines measured by the current-generation ground-based and the hypothesized space-based experiments, respectively. The comparison between true and predicted parameter values demonstrates overall significant constraining power on $\xi$ and $\zeta$, with clear improvement when moving from ground-based to space-based experiments. The errors increase toward higher parameter values, where the power spectra become less sensitive to the steepening of these slopes. The redshift evolution parameter, $\xi_z$, is less constrained compared to $\xi$ and $\zeta$, consistent with the example posteriors shown in figure~\ref{fig:posterior}. Despite the different levels of constraining power on the parameters, the posteriors are shown to be well-calibrated across the parameter space of interest by both the P--P plot and the TARP method, for which the measured validation curves in purple closely follow the ideal expectation shown by the diagonal line in black. We apply the same validation tests to all other cases included in figure~\ref{fig:posterior} and find that similar conclusions hold. This verifies the validity and generalizability of the posteriors inferred by the ILI method. 

\section{Discussion and Conclusions} \label{sec:discussion}

We have presented a flexible ILI framework for constraining the physics of high-$z$ galaxy formation with the summary statistics of multiple LIM signals. Building this ILI framework on simulations generated by \texttt{LIMFAST}, we take the synergy between [$\ion{C}{II}$] and [$\ion{O}{III}$] lines to be measured by forthcoming mm/sub-mm LIM experiments as a case study. We demonstrate that power spectra of these two lines allow us to constrain the physics governing the star formation efficiency and gas depletion timescale of high-$z$ galaxies specified by our power-law parameterization of the mass loading factor and the $\dot{\Sigma}_{\star}$--$\Sigma_\mathrm{g}$ relation. In particular, by adopting the specifications of both forthcoming and hypothesized mm/sub-mm LIM surveys, we show that simultaneous measurements of [$\ion{C}{II}$] and [$\ion{O}{III}$] signals, in either auto-correlation or cross-correlation with galaxies, can effectively lift degeneracies among these physical processes and enable tight, unambiguous constraints on the physical parameters of galaxy formation (e.g., $\xi$ and $\zeta$).  

As our study of galaxy formation parameters suggests, multi-tracer LIM holds strong potential for revealing the physical mechanisms of high-$z$ galaxy formation in the cosmological context. Although we have chosen to focus on the astrophysical parameters by holding the cosmological parameters fixed, a joint analysis of both would be a natural future extension to realize the power of the ILI framework for exploring high-dimensional parameter space. Indeed, substantial degeneracies between astrophysical and cosmological parameters are expected for the case study considered here (Scott \& Sun, in prep.), highlighting the value of varying both sets of parameters and marginalizing over one to obtain unbiased constraints on the other. We note, however, that because astrophysical parameters are typically far more uncertain than cosmological ones, fixing the cosmology to infer astrophysics, as done here, remains a reasonable approach to investigate galaxy formation physics.  

A noteworthy caveat of the ILI framework is its reliance on the assumed forward modeling: if the model is misspecified, the inferred posteriors can remain internally consistent but effectively biased, resulting in misleading or overconfident constraints. Although this is a common limitation for model-based inference, including analytic likelihood approaches, diagnosing it in ILI requires extra care. Since the simple physical model adopted for galaxy formation and line emission will almost certainly be inaccurate in detail, careful comparisons between simulated and observed LIM (and ancillary) summary statistics, along with robustness tests using alternative or perturbed models, are essential for identifying when the framework is significantly biased or being stretched beyond its valid domain. Thus, the reliability of the ILI framework ultimately depends on not only its expressiveness but also the validity and realism of the forward simulations to generate the training data. For example, it will be valuable for future work to incorporate and quantify subtleties that were left out from the present simulations for simplicity, such as redshift-space distortions (RSDs) and the light-cone effect, whose impact on the LIM signals of interest has been shown to be relatively modest but non-negligible by semi-numerical simulations including \texttt{LIMFAST} \citep{LIMFAST1,Murmu2021}. Continued progress in modeling high-$z$ galaxy formation and emission line physics will also enable forward simulations of higher fidelity in light of multi-wavelength, multi-probe observations \citep[][]{Ferrara2019,YangLidz2020,Chakraborty2025,Davies2025,Dhandha2025}. On the other hand, equally important source of potential bias in ILI arises from imperfect modeling of instrumental and survey effects, such as the telescope beam, spectral and spatial response, survey geometry, noise correlations, and foreground mitigation. In practice, oversimplification and/or misspecification of these effects in the forward simulations can systematically bias model inference even if the underlying astrophysical model is robust. Field-level synthetic observations based on end-to-end simulations will therefore be an important goal for future ILI efforts. Finally, the ILI method carries its own potential risks. If the parameter space is not adequately sampled, or if the density estimator (e.g., normalizing flow) is undertrained or extrapolates beyond the support of the training data, the resulting posteriors can be biased or exhibit misleading uncertainties even when the forward model is correct. This highlights again the importance of the validation tests performed in section~\ref{sec:results:validation}. 

While beyond the scope of this paper, several other summary statistics may be of interest and are left to future work for various reasons. The cross-correlation between [$\ion{C}{II}$] and [$\ion{O}{III}$] jointly constrains the production of the two lines and is less susceptible to contamination from line interlopers than auto-correlations. Nevertheless, in addition to causing a large scatter in each individual line luminosity, the diverse and variable star formation histories and ISM conditions of high-$z$ galaxies can lead to a highly stochastic line ratio that de-correlates the two lines even on large scales \citep{Yang2022,Liu2024}. The simple models in this work do not capture these subtleties that can complicate the modeling and interpretation of the cross-power spectrum, although astrophysical scatters can be readily incorporated into the kind of semi-numerical simulations employed here \citep{Reis2022,Nikolic2024}. Meanwhile, lower- and higher-order statistics such as the voxel intensity distribution (VID) and bispectrum are also useful summary statistics to consider for ILI, especially for extracting non-Gaussian information from the line intensity distribution that cannot be captured by the two-point statistics considered in this work. However, the interpretation of these summary statistics can be more challenging due to their higher sensitivity to systematics, more sophisticated covariances, and/or higher computational cost. Finally, it will be valuable for future studies to embed inference frameworks like the one presented here into realistic analysis pipelines that carefully account for instrumental and observational effects, such as survey window functions and foreground mitigation (including the separation of both continuum and interloping lines), that can shape the prospect for probing high-$z$ astrophysics with multi-tracer LIM \cite{FronenbergLiu2024}. 

In summary, our proof-of-concept exploration of the ILI framework in this work represents a promising step toward harnessing the power of multi-tracer LIM to study high-$z$ galaxy formation with the aid of modern machine learning tools. In the future, we will develop further extensions of our modeling and inference methods as well as apply the ILI analysis to a broader range of cosmological probes and datasets.

\section*{Acknowledgments}

We thank the anonymous reviewer for constructive criticism and suggestions for improvements, as well as Matthew Ho, Lun-Jun Liu, Rachel Somerville, and Xiaosheng Zhao for helpful discussions. GS and TN were supported by a CIERA Postdoctoral Fellowship. CAFG was supported by NSF through grants AST-2108230 and AST-2307327; by NASA through grants 21-ATP21-0036 and 23-ATP23-0008; and by STScI through grant JWST-AR-03252.001-A. GS, TN, CAFG, and TS gratefully acknowledge the support of the NSF-Simons AI-Institute for the Sky (SkAI) via grants NSF AST-2421845 and Simons Foundation MPS-AI-00010513. TS was supported by NSF through grant AST-2510183 and by NASA through grants 22-ROMAN22-0055 and 22-ROMAN22-0013. BRS is supported by the LSST Discovery Alliance Data Science Fellowship Program, which is funded by LSST-DA, the Brinson Foundation, the WoodNext Foundation, and the Research Corporation for Science Advancement Foundation. TCC acknowledges support by NASA ROSES grant 21-ADAP21-0122. Part of this work was done at Jet Propulsion Laboratory, California Institute of Technology, under a contract with the National Aeronautics and Space Administration. SRF was supported by NASA through award 80NSSC22K0818 and by the National Science Foundation through award AST-2205900. Part of the analysis was done using the Quest computing cluster at Northwestern University.

\appendix

\section{Fisher Matrix Analysis} \label{sec:fisher}

\begin{figure*}[!ht]
 \centering

 \includegraphics[width=0.49\textwidth]{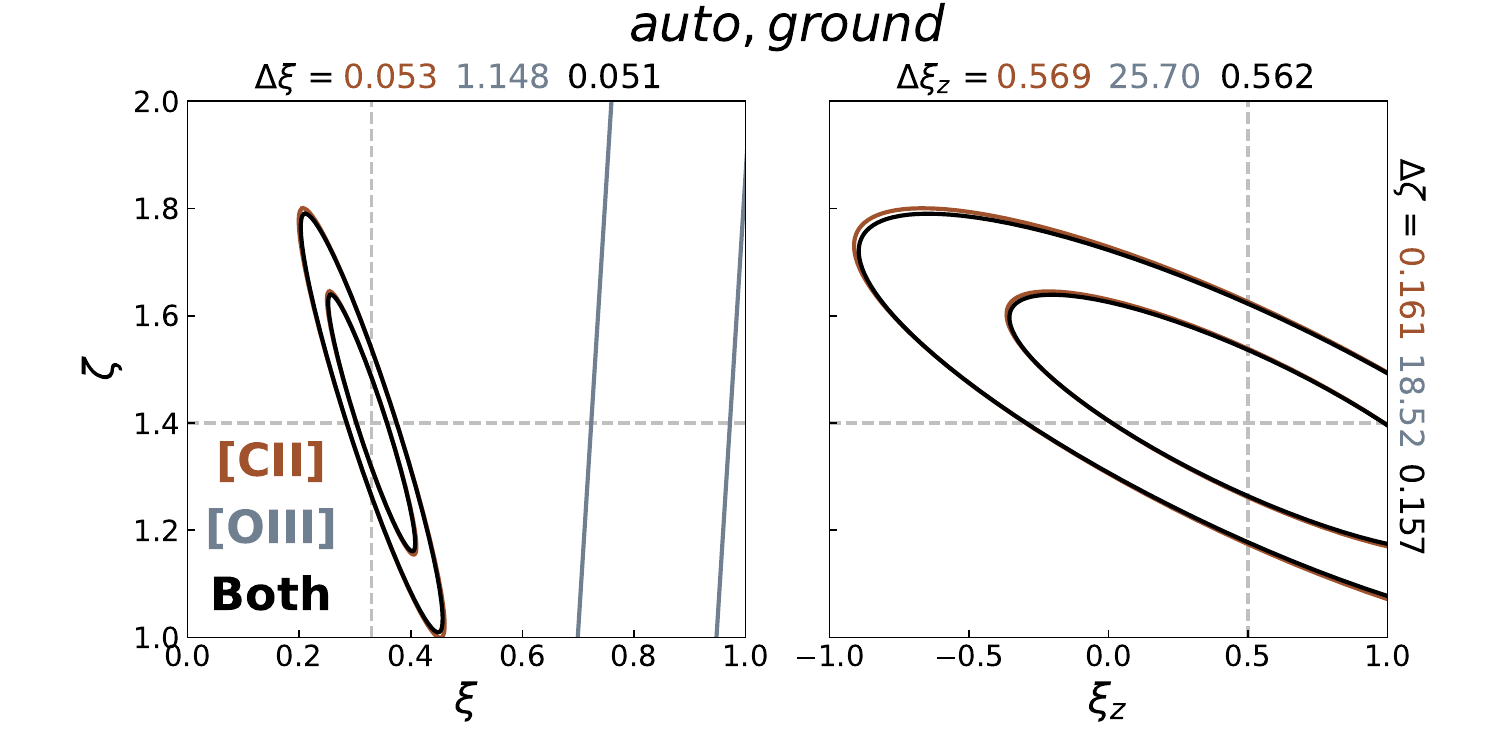}
 \includegraphics[width=0.49\textwidth]{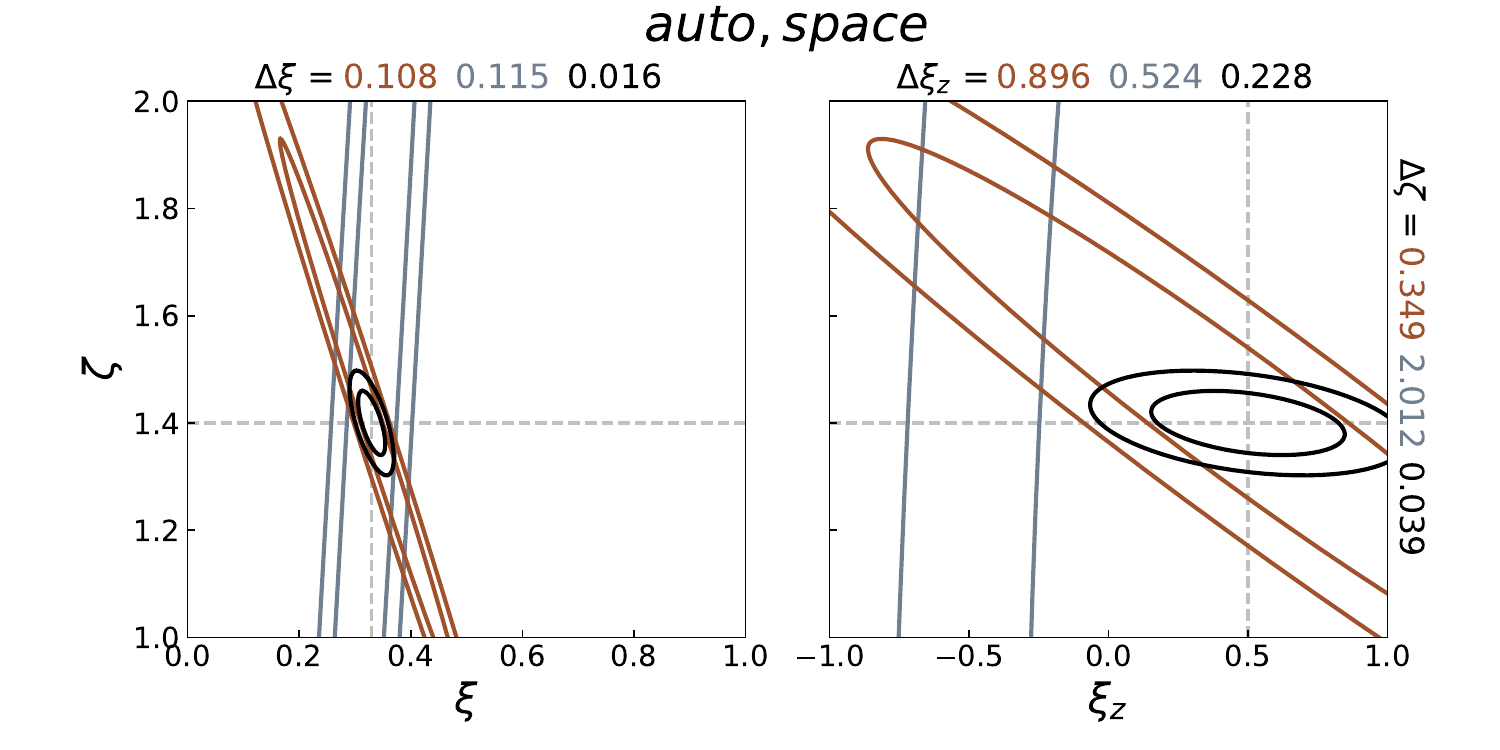}
 \includegraphics[width=0.49\textwidth]{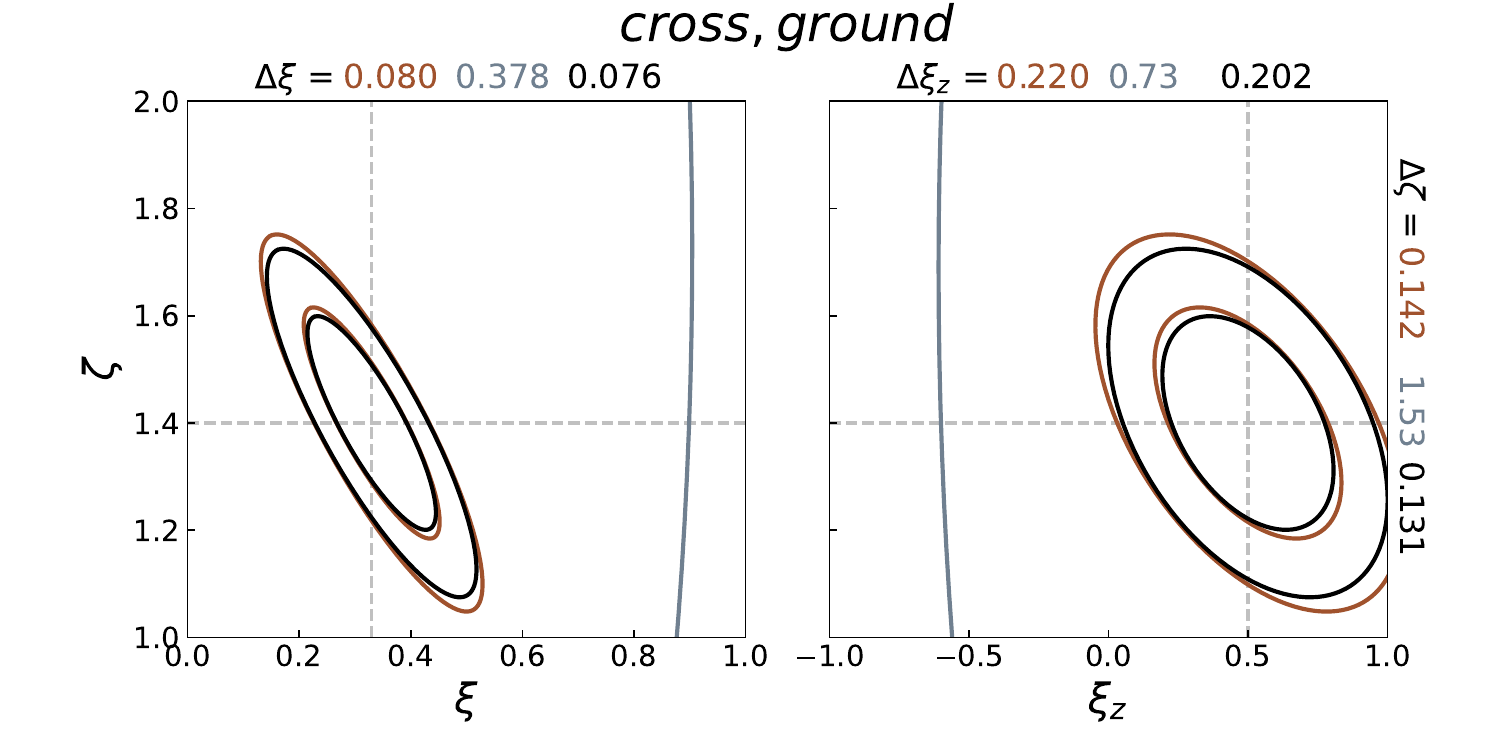}
 \includegraphics[width=0.49\textwidth]{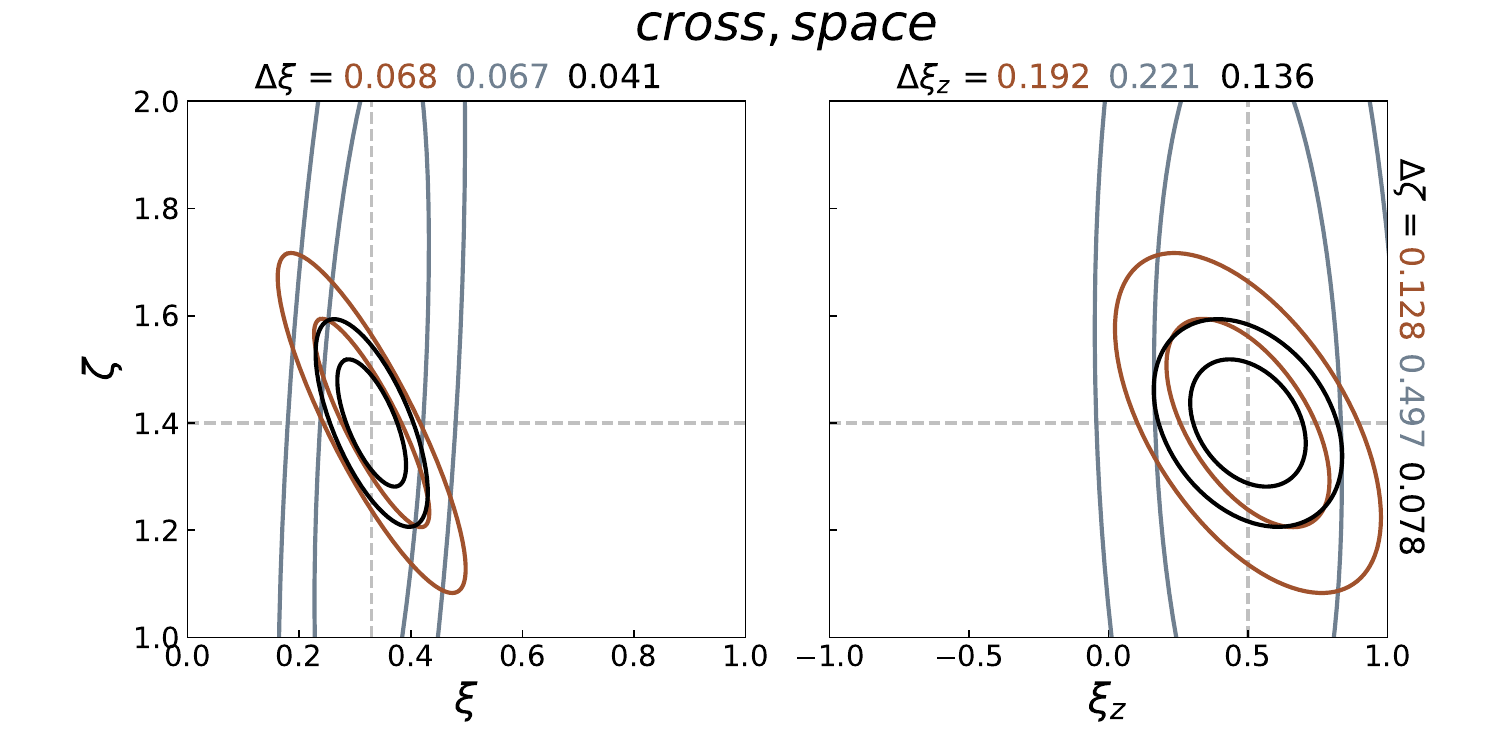}

 \caption{Top: parameter constraints from auto-power spectra predicted by the Fisher matrix analysis for the ground- (left) and space-based (right) surveys, respectively. The inner and outer ellipses represent the 68\% and 95\% confidence levels, respectively, with the marginalized 1$\sigma$ constraints annotated along the axes. Bottom: the same as the top row, but showing instead the constraints from the cross-correlation with Roman LBGs for the ground- and space-based surveys. Note how the combination of both $[\ion{C}{II}]$ and $[\ion{O}{III}]$ lines helps reduce the parameter degeneracies and significantly tighten the constraints.}
 \label{fig:fisher}
\end{figure*}

To better understand the joint posterior distribution obtained from our ILI analysis, it is informative to examine features such as the direction of parameter degeneracies and the overall level of the marginalized constraints and to compare them against predictions from the Fisher matrix analysis. Following \cite{Mason2023}, we run simulations by offsetting the parameter values in a reference model ($\xi=0.3$, $\xi_z=0.5$, $\zeta=1.4$) by $\pm 10\%$ one at a time to evaluate the Fisher matrix
\begin{equation}
F_{ij} = -\frac{\partial^2 \ln \mathcal{L}}{\partial \theta_{i} \partial \theta_{j}} = \frac{\partial^2 \chi^2}{\partial \theta_{i} \partial \theta_{j}} = \sum_{L} \sum_{i_k, i_j} \frac{\partial \Delta_{L}^2(k,z)}{\partial \theta_{i}} \frac{\partial \Delta_{L}^2(k,z)}{\partial \theta_{j}} \frac{1}{\mathrm{var}\left[\Delta_{L}^2 (k,z) \right]},
\end{equation}
where we assume measurements in different $k$ and $z$ bins are independent for a given line $L$. As in our ILI analysis, we calculate the covariance matrix of $\boldsymbol{\theta}$, $\mathbf{C} = \mathbf{F}^{-1}$, from the Fisher matrix for different scenarios with varying summary statistics and survey specifications. We show in figure~\ref{fig:fisher} the parameter constraints based on the Fisher matrix in each of the four cases, where either auto- or cross-power spectra are used under the assumption of the ground- or space-based survey specifications. The Fisher forecasts lead to qualitatively similar conclusions as the ILI-based posteriors, though the latter reveal cases where the distribution is significantly non-Gaussian, beyond the assumption inherent to Fisher analysis. Notably, consistent with the ILI results, the correlation coefficient between $\xi$ and $\zeta$ in the space-based auto-correlation case shifts from $-0.99$ to $-0.67$ when [$\ion{O}{III}$] measurements are included alongside [$\ion{C}{II}$]. 

\begin{table*}
\centering
\caption{Hyperparameters used for training the NPE model (optimized by Optuna).}
\vspace{0.1cm}
\label{tb:hps}
\resizebox{0.8\textwidth}{!}{%
\begin{tabular}{ccccccccc}
\toprule
\textbf{Signal} & \textbf{Observable} & \textbf{Survey} & \textbf{Embedding} & \textbf{Flow} & \textbf{Optimization} \\
 &  &  & $(d_{s}, n_\mathrm{layer})$ & $(n_\mathcal{T})$ & $(\mathrm{LR},\mathrm{WD})$ \\
\midrule
$[\ion{C}{II}]$ & auto & ground & 58, 4 & 4 & $9.5\times10^{-4}, 1.1\times10^{-4}$ \\
$[\ion{O}{III}]$ & auto & ground & 62, 3 & 3 & $8.6\times10^{-4}, 2.3\times10^{-5}$ \\
$[\ion{C}{II}]$ \& $[\ion{O}{III}]$ & auto & ground & 55, 3 & 7 & $7.3\times10^{-4}, 2.7\times10^{-4}$ \\
\midrule
$[\ion{C}{II}]$ & auto & space & 21, 4 & 7 & $7.3\times10^{-4}, 1.1\times10^{-4}$ \\
$[\ion{O}{III}]$ & auto & space & 54, 2 & 2 & $9.5\times10^{-4}, 3.5\times10^{-6}$ \\
$[\ion{C}{II}]$ \& $[\ion{O}{III}]$ & auto & space & 57, 2 & 4 & $7.9\times10^{-4}, 1.7\times10^{-5}$ \\
\midrule
$[\ion{C}{II}]$ & cross & ground & 11, 3 & 7 & $9.1\times10^{-4}, 1.9\times10^{-6}$ \\
$[\ion{O}{III}]$ & cross & ground & 49, 1 & 8 & $8.6\times10^{-4}, 6.1\times10^{-5}$ \\
$[\ion{C}{II}]$ \& $[\ion{O}{III}]$ & cross & ground & 39, 3 & 7 & $6.9\times10^{-4}, 1.5\times10^{-6}$ \\
\midrule
$[\ion{C}{II}]$ & cross & space & 47, 4 & 8 & $8.5\times10^{-4}, 7.4\times10^{-4}$ \\
$[\ion{O}{III}]$ & cross & space & 8, 3 & 8 & $8.2\times10^{-4}, 1.1\times10^{-4}$ \\
$[\ion{C}{II}]$ \& $[\ion{O}{III}]$ & cross & space & 42, 2 & 3 & $9.1\times10^{-4}, 4.9\times10^{-5}$ \\
\bottomrule
\end{tabular}%
}
\end{table*}

\section{Training and Optimization of the NPE Model} \label{sec:ml}

For our NPE model, a Gated Recurrent Unit (GRU; \cite{Cho2014_GRU}) network is adopted to encode sequential inputs of multi-scale and multi-epoch power spectrum data (see figures~\ref{fig:visualize_aps} and \ref{fig:visualize_cps}) into a fixed-dimensional representation of their structures, which is then used to condition a neural spline flow (NSF; \cite{Durkan2019_NSF}) with monotonic rational-quadratic spline transformations that enables the NPE model to flexibly adapt the shape of the inferred posterior distribution to the input features. During training, the flow parameters $\phi$ are optimized by minimizing the negative log-likelihood of the flow (the loss function)
\begin{equation}
    \mathcal{L} = - \mathbb{E}_{(\boldsymbol{\theta}, \boldsymbol{x}) \sim p(\boldsymbol{\theta})p(\boldsymbol{x} \mid \boldsymbol{\theta})} \left[ \log \hat{p}_\phi(\boldsymbol{\theta} \mid \boldsymbol{x}) \right],
\end{equation}
where $\boldsymbol{x}$ denotes the compressed data and $\boldsymbol{\theta}$ the true parameters in the approximate posterior distribution $\hat{p}_\phi(\boldsymbol{\theta} \mid \boldsymbol{x})$. We carry out this minimization using the AdamW optimizer \citep{kingma2014adam,adamw2019}, following a cosine annealing schedule \citep{SGDR} for the learning rate. Training is subject to early stopping based on validation loss to prevent overfitting for generalization performance. Hyperparameters are jointly optimized using Optuna, an optimization framework for hyperparameter tuning \citep{Optuna2019}. Hyperparameter optimization is performed separately for each ILI analysis, tailored to the specific tracer(s) and survey specifications considered, while using consistent optimization ranges and settings for the hyperparameters. Table~\ref{tb:hps} shows a selected number of hyperparameters optimized by Optuna adopted for training, including the dimension of context features ($d_{s}$) and number of layers ($n_\mathrm{layers}$) for the GRU network, the number of transforms ($n_\mathcal{T}$) for the normalizing flow, along with the learning rate (LR) and weight decay (DW) for the optimizer that primarily determine the training performance. 

\begin{figure}[!ht]
 \centering
 \includegraphics[width=0.75\columnwidth]{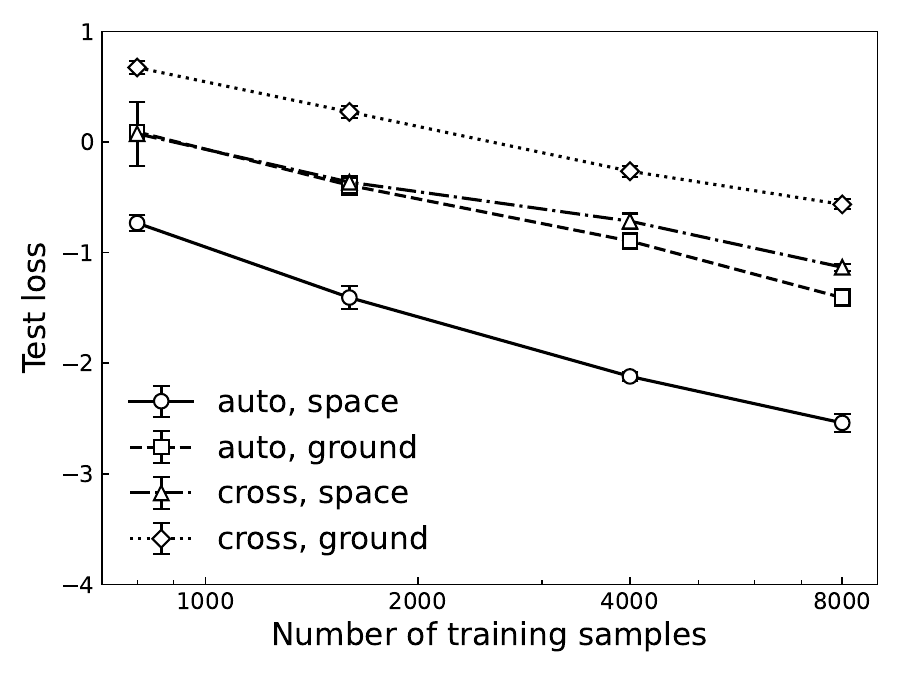}
 \caption{The scaling relation between test loss and the training sample size estimated from training the same NPE model using 10\%, 20\%, 50\%, and 100\% of the designated training set (8000 samples in total). Error bars indicate the scatter from five independent trainings for each sample size.}
 \label{fig:scaling}
\end{figure}

Finally, in figure~\ref{fig:scaling} we characterize how the predictive performance of the NPE model scales with the size of the training set for the four survey configurations considered (auto vs. cross, space vs. ground). Overall, the test loss decreases systematically as the number of training samples is increased from 10\% to 100\% of the total designated training set, following roughly a power law with no clear sign of saturation (i.e., flattening). This indicates that the training is effective but still sample-limited in general. Of the four cases, the ``auto, space'' configuration always shows the largest improvement and achieves the lowest overall loss, consistent with the comparison of parameter constraints shown in figures~\ref{fig:posterior} and \ref{fig:fisher}.

\bibliographystyle{JHEP}
\bibliography{ili.bib}



\end{document}